\documentclass[%
 reprint,
superscriptaddress,
 amsmath,amssymb,
 aps,
raggedbottom,
floatfix,
]{revtex4-2}

\usepackage{graphicx}
\usepackage{dcolumn}
\newcolumntype{+}{D{+}{\,\pm\,}{9,9}}
\newcolumntype{d}[1]{D{.}{.}{#1}}
\usepackage{bm}

\usepackage{enumitem}
\usepackage{array}
\newcolumntype{P}[1]{>{\centering\arraybackslash}p{#1}}

\begin{document}

\preprint{APS/123-QED}

\title{Setting bounds on two-photon absorption cross-sections in common fluorophores with
entangled photon pair excitation}

\author{Kristen M. Parzuchowski}
\affiliation{Department of Physics, 390 UCB, University of Colorado, Boulder, CO, 80309, USA}
\affiliation{JILA, 440 UCB, University of Colorado, Boulder, CO 80309, USA}
\author{Alexander Mikhaylov}%
\affiliation{JILA, 440 UCB, University of Colorado, Boulder, CO 80309, USA}
\author{Michael D. Mazurek}
\affiliation{Department of Physics, 390 UCB, University of Colorado, Boulder, CO, 80309, USA}
\affiliation{National Institute of Standards and Technology, 325 Broadway, Boulder, CO 80305, USA}
\author{Ryan N. Wilson}
\affiliation{Department of Physics, 390 UCB, University of Colorado, Boulder, CO, 80309, USA}
\affiliation{JILA, 440 UCB, University of Colorado, Boulder, CO 80309, USA}
\author{Daniel J. Lum}
\affiliation{National Institute of Standards and Technology, Gaithersburg, MD 20899, USA}%
\author{Thomas Gerrits}
\affiliation{National Institute of Standards and Technology, Gaithersburg, MD 20899, USA}%
\author{Charles H. Camp Jr.}
\affiliation{National Institute of Standards and Technology, Gaithersburg, MD 20899, USA}%
\author{Martin J. Stevens}
\affiliation{National Institute of Standards and Technology, 325 Broadway, Boulder, CO 80305, USA}
\author{Ralph Jimenez}
\email{rjimenez@jila.colorado.edu}
\affiliation{JILA, 440 UCB, University of Colorado, Boulder, CO 80309, USA}
\affiliation{Department of Chemistry, 215 UCB, University of Colorado, Boulder, CO 80309, USA}

\begin{abstract}
 Excitation with entangled photon pairs may lead to an increase in the efficiency of two-photon absorption at low photon flux. The corresponding process, entangled two-photon absorption (E2PA), has been investigated in numerous theoretical and experimental studies. However, significant ambiguity and inconsistency remain in the literature about the absolute values of E2PA cross-sections. Here, we use a fluorescence-based registration scheme to experimentally determine upper bounds on the cross-sections for six fluorophores. These bounds are up to four orders of magnitude lower than the smallest published cross-section. For two samples that have been studied by others, Rhodamine 6G and 9R-S, we measure upper bounds four and five orders of magnitude lower than the previously reported cross-sections. 
\end{abstract}

\maketitle


\section{Introduction}

Two-photon excitation microscopy is a widely used technique for cellular imaging deep within biological tissues. It relies on two-photon absorption (2PA) in a fluorescent molecule, where a nearly simultaneous absorption of two photons leads to a transition from the ground state to an excited state and subsequent fluorescence. Under coherent laser excitation, 2PA is an incredibly unlikely process. To increase the probability of 2PA, light is typically concentrated into short optical pulses and focused to a small spot size, ensuring the photons are well overlapped in time and space~\cite{Denk1990,Zipfel2003}. Nevertheless, most of the incident photons are not involved in 2PA. In biological samples, many of these extra photons instead lead to heating and other forms of damage, which can disrupt biological processes~\cite{Podgorski2016}. 

These practical concerns, along with fundamental interest in quantum metrology and spectroscopy, have stimulated theoretical and experimental studies investigating the possibility of enhancing the efficiency of 2PA by exciting with nonclassical light~\cite{Dorfman2016,Gilles&Knight}. Photon pairs that are entangled in the energy-time and position-momentum degrees of freedom can exhibit the strong temporal and spatial correlations needed for 2PA. Theoretical studies on simple model systems~\cite{Javanainen1990,Gea-Banacloche1989,Fei1997} have predicted that using entangled photon pairs can lead to a significant “quantum advantage” in 2PA rates. Here we define quantum advantage as the ratio of minimum photon flux necessary to observe classical 2PA (C2PA) to that for entangled 2PA (E2PA).

A number of experimental studies have investigated E2PA and concluded that a large quantum advantage does indeed exist~\cite{Lee2006,Upton,Varnavski,Harpham2009,Guzman2010,Villabona-Monsalve2018,Eshun,Villabona-Monsalve2017,Tabakaev,Varnavski2020,Villabona-Monsalve2020}. Some reports suggest the quantum advantage is nearly 10 orders of magnitude~\cite{Harpham2009,Villabona-Monsalve2018}. However, in many of these reports it is unclear whether the signals are caused by E2PA or some other process. For example, some of the reports from Goodson and coworkers~\cite{Lee2006,Upton,Varnavski,Harpham2009,Guzman2010,Villabona-Monsalve2018,Eshun,Varnavski2020,Villabona-Monsalve2020} and Villabona-Monsalve \textit{et al.}~\cite{Villabona-Monsalve2017} conclude that the observed signals are E2PA based only on their linear dependence on photon flux. However, in some cases, the trend of the data and the magnitude of the measurement uncertainty does not exclude a nonlinear fit. Furthermore, a linear dependence is consistent with many one-photon processes such as scattering, one-photon absorption (1PA), or fluorescence from the coating of an optic in the beam path. In Ref.~\cite{Villabona-Monsalve2020}, the potential contributions of several one-photon processes are estimated to be too small to account for the measured E2PA signal. However, this analysis cannot exclude every possible spurious signal. In another study~\cite{Eshun}, the E2PA interaction strength of a particular molecule that was obtained from transmittance measurements is correlated with that obtained from fluorescence measurements. These correlations are consistent with, but not proof of E2PA.

Another signature of E2PA is the signal's dependence on the time delay between photons in a pair; as the time delay is scanned away from optimal overlap (zero time delay), the signal is expected to decrease towards zero, in accordance with the simultaneity requirement of 2PA. In a recent report, Tabakaev \textit{et al.}~\cite{Tabakaev} observe such behavior in an entangled two-photon excited fluorescence (E2PEF) experiment. An interferometer was used before the sample to probabilistically and equally split photons and time delay half of the photon pairs, while the other half traversed the same path. The resulting E2PEF as a function of the time delay should consist of a constant signal due to the photon pairs that traveled the same path and a variable signal due to the photon pairs that traveled different paths. The signal at long time delays should be half the signal at zero delay. Instead, Tabakaev \textit{et al.} observe a signal that tends to zero at large time delays. This unexpected result is pointed out by the authors of the study, but it has yet to be explained. Some of the other reports~\cite{Lee2006,Varnavski,Varnavski2020} have included time-delay scans, but with data reported at only a few delays. Recently~\cite{Mikhaylov2020}, we detailed the difficulty in identifying E2PA signals through transmittance-based schemes with and without implementing a time delay. We compared the time-delay method with a method used previously to measure E2PA that relies on comparing the transmission of photon pairs through a cuvette before and after two-photon absorbers are added~\cite{Upton}. Using the previous technique, we showed that one-photon losses could not be unambiguously distinguished from E2PA. Furthermore, we could reproduce the result of a previous report~\cite{Upton} within a factor of 10 by erroneously attributing the measured one-photon losses to E2PA. In Ref.~\cite{Mikhaylov2020}, we were unable to distinguish an E2PA signal for Zinc tetraphenylporphyrin from the noise, however the time-delay technique enabled us to bound the maximum efficiency of E2PA for this particular molecule to be more than 100 times lower than anticipated based on a previous report~\cite{Upton}. The achieved sensitivity for measuring a change in transmittance ($\approx 1 \%$) was limited by residual interference resulting from the time-delay scheme. Variations on this time-delay scheme that implement improved fringe averaging techniques can increase the sensitivity achieved in this pilot study. In nearly all the aforementioned reports by other groups, the E2PA cross-section sensitivity---and thus the measurement confidence level---is left unspecified. These ambiguities leave unanswered questions regarding the magnitude of the quantum advantage.

Although the tantalizing prospect of a large quantum advantage remains, experimental methodology has not yet evolved to a point where meaningful comparisons between experiments by different groups is straightforward, where interpretation in theoretical context is possible, or even where the relative magnitudes of the C2PA and E2PA signals can be measured in the same experiment. Our study exposes and addresses some of these issues. We emphasize the role of the spatio-temporal correlations of the excitation source on E2PA and the importance of providing these characteristics when reporting the absolute cross-section values. 

To directly address the question of the quantum advantage, we present a method for measuring both E2PEF and classical two-photon excited fluorescence (C2PEF) in one experimental setup. We characterize our excitation sources, fluorescence collection system and samples to determine our 2PA cross-section sensitivity. The C2PEF measurements are used to derive C2PA cross-sections, $\sigma_{C}$, for the six studied fluorophores. The values strongly agree with previously reported C2PA cross-sections~\cite{deReguardati2016,Eshun,Meiling2018}. Although we do not detect measurable E2PEF signals for any of the six fluorophores, we can bound the maximum efficiency of the E2PA process in each fluorophore by placing upper bounds on its E2PA cross-section, $\sigma_E$ (defined more precisely in Section~\ref{theory}). The C2PEF and E2PEF measurements are also used to bound the quantum advantage. Our established upper bounds on $\sigma_E$ are up to four orders of magnitude lower than the smallest published value of $\sigma_E$~\cite{Tabakaev}. For two of the samples, the upper bounds on $\sigma_E$ are four and five orders of magnitude lower than the previously reported cross-sections~\cite{Tabakaev,Eshun}. 

In Section~\ref{theory}, we present a connection between a simple, probabilistic theory describing E2PA and the well-accepted description of C2PA. In Section~\ref{LitSum}, we provide a literature summary of selected E2PA experimental results that can be compared to the results of our experiment. Afterwards, we discuss our experimental setup and the characterization of the entangled photon source and fluorescence collection unit in Section~\ref{Exp}. We present our results---including the upper bounds on the E2PA cross-section and quantum advantage---in Section~\ref{results}. Further details about the experiment and analysis are given in the appendices.

\section{Theoretical background}
\label{theory}
The rates of multi-photon processes are sensitive to the photon statistics~\cite{Mollow1968,Spasibko2017}, which can be characterized by the second-order coherence, $g^{(2)}=\left\langle \left.\hat{a}^\dagger\right.^2\hat{a}^2\right\rangle\Big/\left\langle \hat{a}^\dagger\hat{a}\right\rangle^2$, where $\hat{a}^\dagger$ and $\hat{a}$ are the photon creation and annihilation operators. For a single-mode field with mean photon number $\mu = \left\langle \hat{a}^\dagger\hat{a}\right\rangle$, the 2PA rate can be written~\cite{Gilles&Knight,Weber1971}
\begin{equation}
\label{Rg2}
     R = \kappa_2 \left\langle \left.\hat{a}^\dagger\right.^2\hat{a}^2\right\rangle = \kappa_2 \mu^2 g^{(2)},
\end{equation}
where $\kappa_2$ (s$^{-1}$) is a collection of constants quantifying the strength of the nonlinear interaction. It has been demonstrated, for example, that thermal light ($ g^{(2)}=2$) doubles the 2PA rate compared to laser (coherent) excitation ($g^{(2)}=1$) of the same intensity~\cite{Jechow2013}.

In the classical limit, the instantaneous 2PA rate for a single fluorophore can be written~\cite{Rumi2010}
\begin{equation}
\label{C2PArate}
     R = \frac{1}{2}\sigma_C \phi^2,
\end{equation}
where $\phi$ is the photon flux, with units of cm$^{-2}$~s$^{-1}$. The C2PA cross-section $\sigma_C$ has units of GM, where $1\,\mathrm{GM} = 10^{-50}$~cm$^4$~s~. (In this section we omit photons, excitations and fluorophores from the units of various quantities for brevity; in later sections we include them for clarity.)

For a pulsed laser ($g^{(2)}=1$) source with temporal and spatial mode set by the pulse duration $T$ (fs) and the beam area $A$ (cm$^2$), if we  rewrite Eq.~\eqref{Rg2} in terms of the photon flux, $\phi=\mu/(T A)$, and substitute
\begin{equation}
\label{kappa_2}
    \kappa_2 = \frac{\sigma_C}{2 T^2 A^2},
\end{equation}
we arrive at the classical limit in Eq.~\eqref{C2PArate}.

In contrast to laser light, spontaneous parametric downconversion (SPDC) produces photon pairs exhibiting correlations in energy, time and space that can be tailored to enhance the rate and selectivity of 2PA~\cite{Schlawin2018,Leon-Montiel2019}. The energy correlations between the signal and idler photons within a pair are set by conservation of energy in the conversion of one pump photon to two down-converted photons and can be engineered to match the energy of a two-photon transition. Photon pair production is localized in space and time~\cite{Schneeloch2016,Boyd2018}, allowing for excitation with photons that nearly simultaneously arrive in a localized region of space.

A degenerate, single-mode~\footnote{Signal and idler photons occupy a single optical mode in all degrees of freedom (polarization, spatial mode, spectral and temporal profiles)} SPDC source can be modeled as a single-mode squeezed vacuum (SMSV), for which $g^{(2)} = 3+1/\mu$~\cite{Gilles&Knight}. Substituting this expression into Eq.~\eqref{Rg2} yields
\begin{equation}
\label{Rsmsv}
R =  \kappa_2 \left(\mu + 3\mu^2\right).
\end{equation}
For a pulsed source, substituting $\phi$ and Eq.~\eqref{kappa_2} into Eq.~\eqref{Rsmsv} gives
\begin{equation}
R =  \frac{1}{2}\sigma_C \left(\frac{\phi}{T A} + 3\phi^2 \right). 
\end{equation}
An alternative way to write Eq.~\eqref{Rsmsv} is~\cite{Fei1997}
\begin{equation}
\label{E2PArate}
R_E =  \frac{1}{2}\left(\sigma_E \phi +3 \sigma_C \phi^2\right),
\end{equation}
where the E2PA cross-section has units of cm$^2$~\footnote{We follow the convention that the 2PA cross-section refers to the removal rate of two photons from the field for each excited molecule; thus is it different by a factor of two from that in Ref.~\cite{Fei1997}} and $R_E$ is the instantaneous E2PA rate. For a single-mode field, the two cross-sections are related by $\sigma_{\mathrm{E}} = \sigma_C / (T A)$. At low photon flux, the first term dominates and the E2PA process should scale linearly with $\phi$~\cite{Javanainen1990,Gea-Banacloche1989}~\footnote{More precisely, the E2PA rate is linear in \textit{photon number}, as Eq.~\eqref{Rsmsv} shows. If pulse duration or spot size were modified, the linear term should scale in the same way as the quadratic term, at least in the single-mode case considered here.}. At high photon flux, where many photon pairs overlap in time, the quadratic term dominates.

\begin{figure}[ht]
\includegraphics[width=0.48\textwidth]{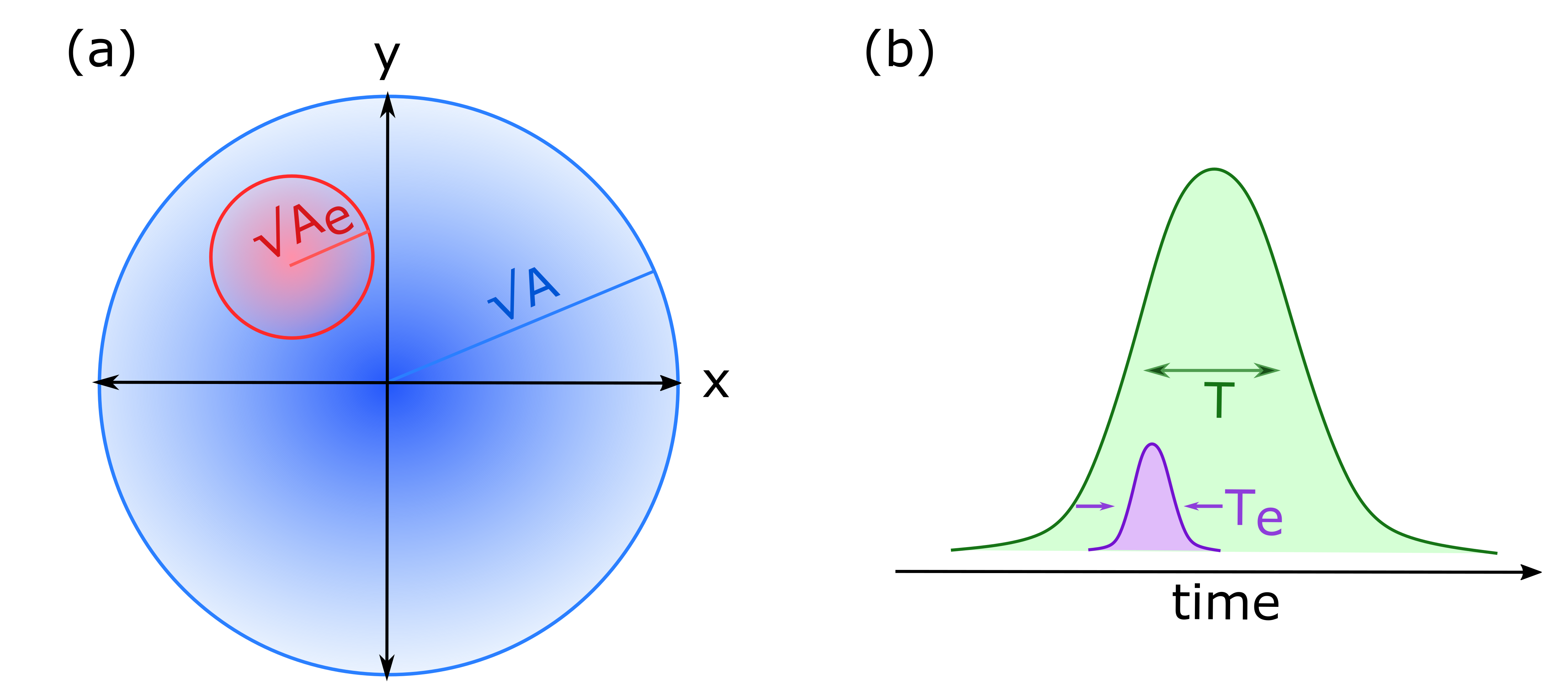}
\caption{\label{fig:AeTeScheme}(a) The width of the 2-dimensional distribution of an idler photon's position conditioned on the signal photon's position is $\propto \sqrt{A_e}$ where $A_e$ is the entanglement area. Photons closely correlated in space may have $A_e<<A$ where $A$ is the SPDC beam area. (b) The entanglement time, $T_e$, is the width of the distribution of idler arrival times conditioned on signal arrival time. Photons closely correlated in time may have $T_e$ smaller than the duration of either the pump pulse or the overall SPDC pulse, $T$.}
\end{figure}

In a real experiment, the SPDC light typically occupies multiple modes and Eq.~\eqref{Rsmsv} does not hold. In this case, the coefficient for the linear term could in principle be larger than the coefficient for the quadratic term. Roughly speaking, if the two photons in a pair are more closely correlated in time than the pump pulse duration, the interaction strength could be enhanced by the factor $T/T_{\mathrm{e}}$, where $T_{\mathrm{e}}$ is the entanglement time. Analogously, if the photons in a pair are more closely correlated in space than the beam size, the interaction strength could be modified by the factor $A/A_{\mathrm{e}}$, where $A_{\mathrm{e}}$ is the entanglement area. The values of $T_{\mathrm{e}}$ and $A_{\mathrm{e}}$ are set by the temporal and spatial $g^{(2)}$ functions~\cite{Fei1997,Jost1998} and schematically illustrated in Fig.~\ref{fig:AeTeScheme}. Following this simple, probabilistic argument leads to the approximation~\footnote{Other than a factor of two difference due to our different definition of $\sigma_E$, this is the same approximation arrived at in Ref.~\cite{Fei1997}, but following a different argument and making different assumptions.}
\begin{equation}
\label{e2PAtoc2PA}
    \sigma_E \approx \frac{\sigma_C}{T_{\mathrm{e}}A_{\mathrm{e}}}.
\end{equation}
To maximize the E2PA rate, $T_{\mathrm{e}}$ and $A_{\mathrm{e}}$ should be as small as possible. For a large $\mu$ and a large number of modes,  $g^{(2)}\xrightarrow{}1$ and the E2PA rate approaches the classical limit in Eq.~\eqref{C2PArate}.

Unlike $\sigma_{C}$, which depends only on wavelength for a particular molecular 2PA transition, the value of $\sigma_{E}$ depends strongly on the properties of the excitation source and experiment when defined this way. The values of $A_e$ and $T_e$ evolve as the SPDC beam propagates through optics from the downconversion crystal to the sample~\cite{Valencia2002,Pittman1996,Zhang2019,Edgar2012,Jost1998,Unternahrer2018}, and therefore depend on the details of the optical system used to measure E2PA. Calculating $\sigma_{E}$ for a given experimental geometry thus requires knowledge of $A_e$ and $T_e$ within the excitation volume. Clearly, these factors complicate the ability to compare results from different experiments.

\section{Literature Summary}
\label{LitSum}
\begin{table*}
    \caption{\label{Tab:e2PALiterature} Results and experimental parameters from selected E2PA studies. Cross-sections ($\sigma_{C}$ and $\sigma_{E}$) are quoted at the corresponding excitation wavelength ($\lambda$). Cross-sections and entanglement times ($T_e$) are taken directly from the reports unless otherwise noted. We estimate the entanglement area ($A^{\mathrm{est}}_e$) required to explain the $\sigma_E$ values based on Eq.~\eqref{e2PAtoc2PA}.}
    \begin{ruledtabular}
    \begin{tabular}{lccccc}
    Sample~[Ref.]  & $\lambda$ &  $\sigma_{C}$ & $\sigma_{E}$ & $T_e$ & $A^{\mathrm{est}}_{e}$\\ 
    & (nm)   &  (GM)   & (10$^{-19}$~cm$^2$~fluorophore$^{-1}$) & (fs)  & (10$^{-9}$~$\mu$m$^{2}$) \\ \hline
    9R-S~\cite{Eshun}& 800 & 27.9 & $2.02-2.69$ &  100 & $1.0-1.4$ \\
    Rh6G~\cite{Tabakaev} & 1064 &  $9.9\pm1.5$\footnotemark[1] & $0.0099-0.019$ & 140\footnotemark[2]  &  $38-72$ \\
    RhB~\cite{Villabona-Monsalve2017} & 808  & $260\pm40$\footnotemark[1] & $0.17-42$ & 17  &  $3.6-900$\\
    Tetraannulene~\cite{Guzman2010} & 800 & 2960 & 990 & 96   & 0.31 \\
    \end{tabular}
    \end{ruledtabular}
    \footnotetext[1]{From Ref.~\cite{Makarov2008}}
    \footnotetext[2]{Not explicitly written in report, but we estimate based on reported details}
\end{table*}

A summary of important parameters in selected E2PA reports is given in Table~\ref{Tab:e2PALiterature}. These studies represent experimental efforts from several independent groups. Table~\ref{Tab:e2PALiterature} shows $\sigma_C$ and $\sigma_E$ values determined at several near infrared wavelengths. In Ref.~\cite{Eshun}, pulsed-pumped type-II~\footnote{The two photons in a pair have perpendicular polarizations, rather than the parallel polarizations of type-0 or type-I SPDC. Studies of C2PA~\cite{mcclain1973,lakowicz1996,Rapaport2003} suggest that there is a small difference in efficiency (by a factor of two or three) of excitation when the two photons have perpendicular polarizations instead of parallel.} SPDC was generated to excite a sample with $1-25\times10^6$~photons~s$^{-1}$~\footnote{the beam waist is not specified} in transmittance- and fluorescence-based E2PA schemes. For the studied 9R-S molecule, the $\sigma_E$ values found using these two techniques differ slightly from one another. The measurement uncertainty was estimated to be 9$\%$ and 12$\%$ for transmittance- and fluorescence-based techniques. In Ref.~\cite{Tabakaev}, a continuous-wave (CW)-pumped~\footnote{The linear photon-flux-dependent term in Eq.~\eqref{E2PArate} will remain dominant over the quadratic term at higher average fluxes for a CW-pumped SPDC excitation source than for that of a pulsed-pumped SPDC excitation source. This is because of the lower likelihood of uncorrelated pairs arriving at a fluorophore at the same time.} type-0 SPDC source was used for E2PEF measurements with an effective incident photon rate of $2-50\times10^7$~photons~s$^{-1}$ (beam waist of 60~$\mu$m). A 100 times increase of the molar concentration led to a decrease in the measured $\sigma_E$ value for Rhodamine 6G (Rh6G) by a factor of two. The uncertainties on the measured cross-sections were estimated to be nearly 50$\%$. A similar concentration dependence was observed in Ref.~\cite{Villabona-Monsalve2017} using pulsed-pumped type-II SPDC excitation with an incident pair rate of $50-7,000$~photon~pairs~s$^{-1}$ (beam waist of 61~$\mu$m) in a transmittance-based scheme, where the concentration dependence of $\sigma_E$ for Rhodamine B (RhB) was attributed to potential aggregation effects in the solutions. The uncertainties on the published cross-sections are $\approx10\%$. In Ref.~\cite{Guzman2010}, a pulsed-pumped type-II SPDC source was used to excite the tetraannulene sample with $1-25\times10^6$~photons~s$^{-1}$ in a transmittance-based E2PA scheme. The measurement uncertainty was not estimated in this report. The values for $\sigma_{E}$ of $990\times10^{-19}$~cm$^2$~fluorophore$^{-1}$ for tetraannulene~\cite{Guzman2010} and $0.0099\times10^{-19}$~cm$^2$~fluorophore$^{-1}$ for Rh6G~\cite{Tabakaev} are the largest and smallest values, respectively, that have been reported. In all these reports the photon flux is not specified, except for an order-of-magnitude estimate in Ref.~\cite{Villabona-Monsalve2017}, and the photon rate is not precisely defined.

No direct $T_e$ measurements were completed in the aforementioned reports. In the case of Ref.~\cite{Tabakaev} we estimate $T_e$ based on the details provided by the authors, who estimated an effective flux reduced to the fraction of photon pairs that have $T_e=$~140~fs. In Refs.~\cite{Eshun,Villabona-Monsalve2017,Guzman2010}, the value of $T_e$ was estimated at the output of the crystal, which in some cases can be orders of magnitude smaller than the value at the sample's position. This is especially true when the total group delay dispersion (GDD) of the optics is large or when the bandwidth of the SPDC is large (Appendix~\ref{TimeAndArea}). The value of $T_e$ would be very sensitive to even small amounts of GDD in the later case. 

The values of $A_e$ are not specified in any of these reports. We estimate the entanglement areas, $A_e^{\mathrm{est}}$, required to explain the results of these previous reports, based on the probabilistic model, Eq.~\eqref{e2PAtoc2PA}. The values of $A_e^{\mathrm{est}}$ range from $10^{-6}-10^{-10}$~$\mu\mathrm{m}^2$. This would require both photons within a pair to be confined to a region that is $10^{-6}-10^{-10}$ times the diffraction limited spot size. We have no evidence that this level of confinement is feasible~\cite{Zhang2019,Edgar2012,Pittman1996,Jost1998}. Thus, the simple, probabilistic theory~\cite{Fei1997} used to derive Eq.~\eqref{e2PAtoc2PA} cannot explain these experimental results. More sophisticated theoretical models allowing for quantum interference and other effects might be able to explain some additional enhancement. For example, Ref.~\cite{Burdick2018} predicts an additional enhancement of up to three orders of magnitude over the probabilistic model for the molecule N$_2$. Furthermore, the recent theoretical model used to calculate $\sigma_E$ of thiophene dendrimers in Ref.~\cite{Schatz2020} produces results that are consistent with the experimental results of Ref.~\cite{Harpham2009}, but relies on assumptions about linewidths that have yet to be experimentally verified. In contrast, our recent report~\cite{Mikhaylov2020} sets an upper bound on $\sigma_E$ for Zinc tetraphenylporphyrin,  $\sigma_E^{\textrm{UB}}=1.7\times10^{-19}$~cm$^2$, that is significantly lower than earlier reports had claimed, $\sigma_E=2.37\times10^{-17}$~cm$^2$~\cite{Upton}.


\section{Experimental Setup and Characterization}
\label{Exp}
\begin{figure*}[t]
\includegraphics[width=0.9\textwidth]{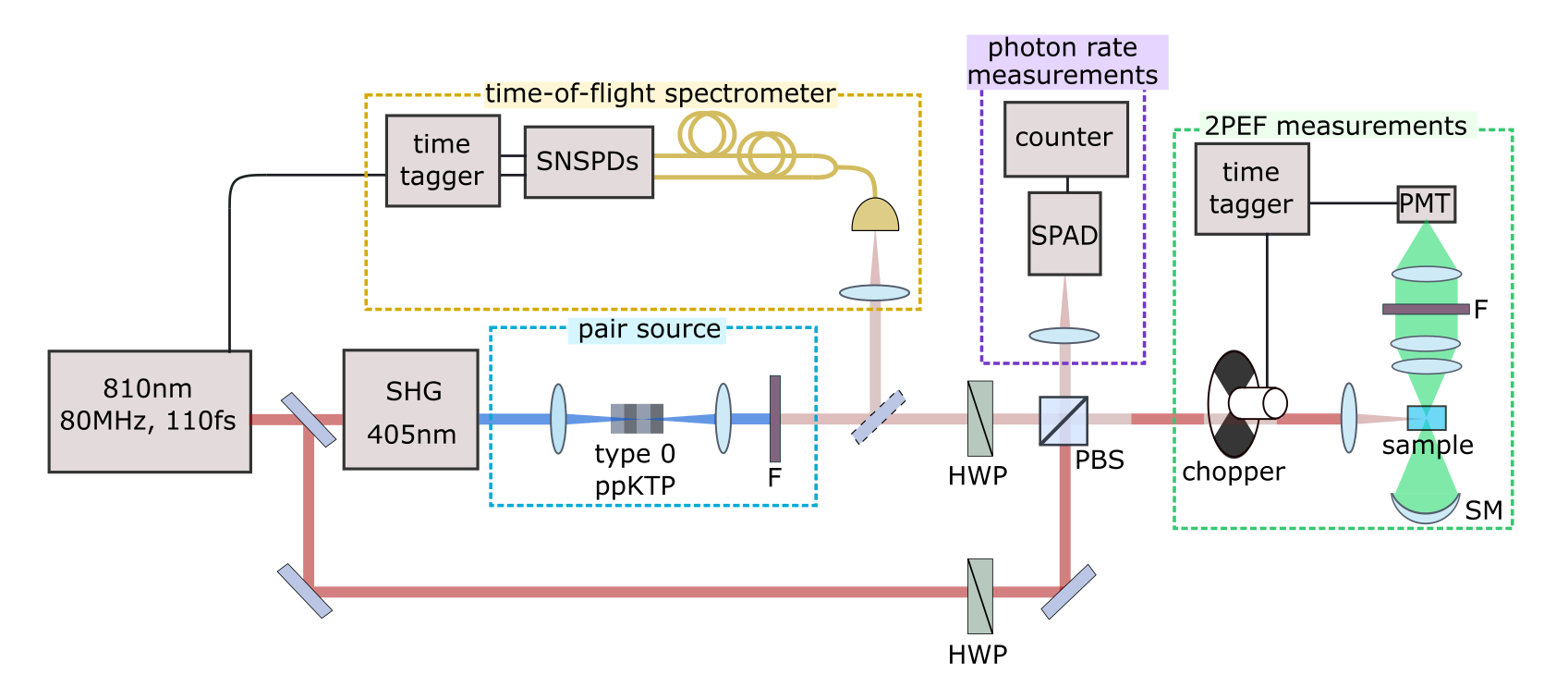}
\caption{\label{fig:setup}Schematic of the experimental setup. The 810~nm laser (dark red) is split into two paths; one path is used for C2PEF measurements and the other for E2PEF measurements. The light in the C2PEF path is directed through a half-wave plate (HWP) and polarizing beam splitter (PBS) to control the power of light directed to the 2PEF measurement system. The laser is optically chopped (chopper) and focused into a sample. The fluorescence (green) is collected onto a photon-counting PMT and all scattered light rejected using filters (F). The PMT pulses are shaped and sent to a time tagger that is synchronized with the optical chopper. The light in the E2PEF path is frequency doubled (blue) via second-harmonic generation (SHG) and focused into a type-0 ppKTP crystal to generate collinear SPDC photon pairs at 810~nm (light red). Filters (F) are used to remove the remaining 405~nm light.  To characterize the joint spectral intensity of the light, a flip mirror directs the pairs into a time-of-flight spectrometer~\cite{Avenhaus2009,Gerrits2015}. To characterize the absolute SPDC photon rate at the sample, a HWP and PBS direct the light to a single-photon avalanche diode (SPAD). For E2PEF measurements, the HWP is rotated to transmit all SPDC photons through the PBS; this light travels along the same path as the light used in the C2PEF measurements where it is focused into a 2PA sample.}
\end{figure*}

Here we give a brief overview of our experimental setup and characterizations. A thorough description of the components is given in Appendix~\ref{ExpDetails}. A schematic of the experimental setup is shown in Fig.~\ref{fig:setup}. A pump laser emits $\approx 110$~fs pulses with a center wavelength of 810~nm ($\approx 9$~nm bandwidth) at a repetition rate of $8\times10^7$~pulses~s$^{-1}$. The laser output is frequency doubled to produce 405~nm light ($\approx 3$~nm bandwidth), of which 30~mW is focused into a type-0 periodically poled potassium titanyl phosphate (ppKTP) crystal to generate SPDC for E2PA. A small fraction of the pump laser output is routed around the nonlinear crystals and used for C2PA.

We characterize the joint spectrum of the photon pairs with a time-of-flight fiber spectrometer consisting of 500~m-long single-mode fibers and superconducting nanowire single-photon detectors (SNSPDs) (Appendix~\ref{TimeAndArea}). The SPDC is approximately degenerate and centered at 810~nm with $\approx 76$~nm bandwidth. We determine the entanglement time (Appendix~\ref{TimeAndArea}) using the estimated joint temporal intensity, which accounts for the approximately $3700\,\mathrm{fs}^2$ of dispersion accumulated by each photon pair before reaching the center of the cuvette. The value of $T_e$ at the sample position is $\approx1620\,\mathrm{fs}$. Although $T_e$ is larger than in the ideal (dispersion-free) case, we make a probabilistic estimate that lossless dispersion compensation would at most increase the rate of E2PA by a factor of 95 (Appendix~\ref{TimeAndArea}). 

The SPDC photon rate is measured using a free space coupled single-photon avalanche diode (SPAD). The optical system was designed to minimize losses, thereby minimizing the number of unpaired photons focused into the sample. Taking into account the single-photon detection rate, the SPAD dead time and efficiency, the photon statistics of the SPDC, and the optical losses in our setup from the center of the crystal to the center of the sample ($\approx24\%$), we estimate that $\approx147$~photons~pulse$^{-1}$ are generated at the output of the crystal and $\approx112$~photons~pulse$^{-1}$ arrive at the sample while operating at our maximum pump power (30~mW) (Appendix~\ref{MeanPhotNum}).

Unfortunately, we do not have a direct measurement of entanglement area; we can only estimate that the value of $A_e$ is in the range of $2.1-13,700\,\mathrm{\mu m}^2$. The estimate of the lower bound is based on the diffraction limit. We find no evidence that the two photons can be focused to a region significantly smaller than that set by the diffraction limit. It has been shown~\cite{Stohr2018,Rozema2014,Unternahrer2018,Edgar2012,Zhang2019,Boyd2018} that entangled photons can be focused to a spot size that is a few-fold smaller, however we neglect these factors here as they have a minor effect in the orders-of-magnitude comparisons we present. Thus, we set the bound using a circular area with radius ($r$) set by the central wavelength of excitation ($r \approx \lambda$). The estimate of the upper bound is set by an elliptical area with diameters set by the measured FWHM of the beam in transverse directions at the center of the sample~\footnote{The value of $A_e$ likely changes throughout the cuvette because the beam is not collimated. The range of values that $A_e$ could take on at the edge of the cuvette is $2.1-2,160,000\,\mu$m$^2$. This subtle point is taken into account in Appendix~\ref{DataAn}}.

\begin{table*}[t]
    \caption{\label{Tab:e2PA limits}Summary of literature C2PA cross-sections ($\sigma^{\mathrm{lit}}_{C}$), measured C2PA cross-sections ($\sigma^{\mathrm{exp}}_{C}$), measured E2PA cross-section upper bounds ($\sigma_{E}^{\mathrm{UB}}$), estimates for the E2PA cross-sections ($\sigma_{E}^{\mathrm{est}}$) (based on Eq.~\eqref{e2PAtoc2PA} using $T_e = 1620\,\mathrm{fs}$ and $A_e = 2.1\,\mathrm{\mu m}^2$), and measured quantum advantage upper bounds (QA$^{\mathrm{UB}}$). All quantities are listed for 810~nm excitation unless otherwise noted.}
    \begin{ruledtabular}
    \begin{tabular}{ l+++c+ }
    Sample  & \multicolumn{1}{c}{$\sigma_{C}^{\mathrm{lit}}$~[Ref.]}& \multicolumn{1}{c}{$\sigma_{C}^{\mathrm{exp}}$}& \multicolumn{1}{c}{$\sigma_{E}^{\mathrm{UB}}~\times~10^{25}$} & \multicolumn{1}{c}{$\sigma_{E}^{\mathrm{est}}~\times~10^{30}$} & \multicolumn{1}{c}{$\mathrm{QA}^{\mathrm{UB}}$}   \\ 
    & \multicolumn{2}{c}{(GM)} & \multicolumn{2}{c}{(cm$^2$~fluorophore$^{-1}$)} &\\ \hline
    AF455 & 350+30\text{~\cite{deReguardati2016}}& 660+180 & 2.1+0.5  &190 & 410+140\\
    Qdot 605 & 27000+8000\footnotemark[1]\text{~\cite{Meiling2018}}& 46000+13000 & 480+120 &14000 & 730+240\\
    Fluorescein & 21+2\text{~\cite{deReguardati2016}}& 13+4 & 1.0+0.2 & 3.8 & 2000+700\\
    9R-S & \multicolumn{1}{c}{27.9\footnotemark[1]\,\cite{Eshun}}& 22+6 & 20+6 & 6.5 & 7000+2000\\
    Rh6G & 78+7\text{~\cite{deReguardati2016}}& 51+14 & 1.2+0.3 & 15 & 1100+400\\
    C153 & 17+2\text{~\cite{deReguardati2016}}& 14+4 & 1.6+0.4 & 4.1 & 2400+800 \\
    \end{tabular}
    \end{ruledtabular}
    \footnotetext[1]{Measured at 800~nm}
\end{table*}

For C2PEF and E2PEF measurements, we use a polarizing beamsplitter (PBS) to combine the SPDC and laser beams, and align them along the same path. The power of the laser beam is controlled using a half-wave plate (HWP) in conjunction with the PBS, varying from $0.079-10.5\,\mu$W. The beams are sent through an optical chopper, then focused in the center of a cuvette to a beam FWHM of $\approx 68\,\mathrm{\mu m}$ and $\approx 49\,\mathrm{\mu m}$ for the SPDC and laser beams respectively. For E2PEF measurements, we block the laser beam, and for C2PEF measurements we block the SPDC beam. The portion of the beam absorbed in the sample is partially re-emitted as fluorescence, which is collected and focused onto a photon-counting photomultiplier tube (PMT). A combination of a shortpass and bandpass filter (selected for each fluorophore, see Appendix~\ref{DataAn}) in front of the PMT reject scattered 810~nm and 405~nm light. The SPDC beam is found to have a larger divergence within the sample compared to the laser beam (Appendix~\ref{Align}). The divergence is taken into account by using the characterization of the spatially dependent geometrical collection efficiency (Appendix~\ref{DataAn}). The geometrical collection efficiency of the fluorescence collection system is characterized using numerical simulations and 1PEF measurements (Appendix~\ref{CE}) and determined to be 15.4$\%$ and 4.7$\%$ for a point source and line source (extending the length of the cuvette) of fluorescence, respectively. The two beams are found to be displaced from each other in the sample by $\approx 5\,\mu$m horizontally and vertically. As Appendices~\ref{Align} and~\ref{CE} explain, our experimental apparatus is carefully designed and characterized to be robust against small changes in alignment like these. The longitudinal displacement between the beams is compensated for. 

The six fluorophores investigated in this study are the 1,3,5-triazine-based octupolar molecule ``AF455"~\cite{Kannan2004,Rogers2004} in toluene, Qdot ITK carboxyl quantum dot 605 (qdot 605) in borate buffer, fluorescein in pH 11 water, the benzodithiophene derivative ``9R-S"~\cite{Eshun} in chloroform, rhodamine 590 (Rh6G) in methanol and coumarin 153 (C153) in toluene (details on sample preparation in Appendix~\ref{samplePrep}). These samples are of particular interest because of their well-known and large values of $\sigma_C$ at 810~nm  (see Table~\ref{Tab:e2PA limits}). In addition, two of these samples (Rh6G and 9R-S) were studied in previous reports of E2PA~\cite{Tabakaev,Eshun}.

\section{Results and Discussion}
\label{results}
\begin{figure*}[p]
\includegraphics[trim=30 40 40 45,clip,width=0.67\textwidth]{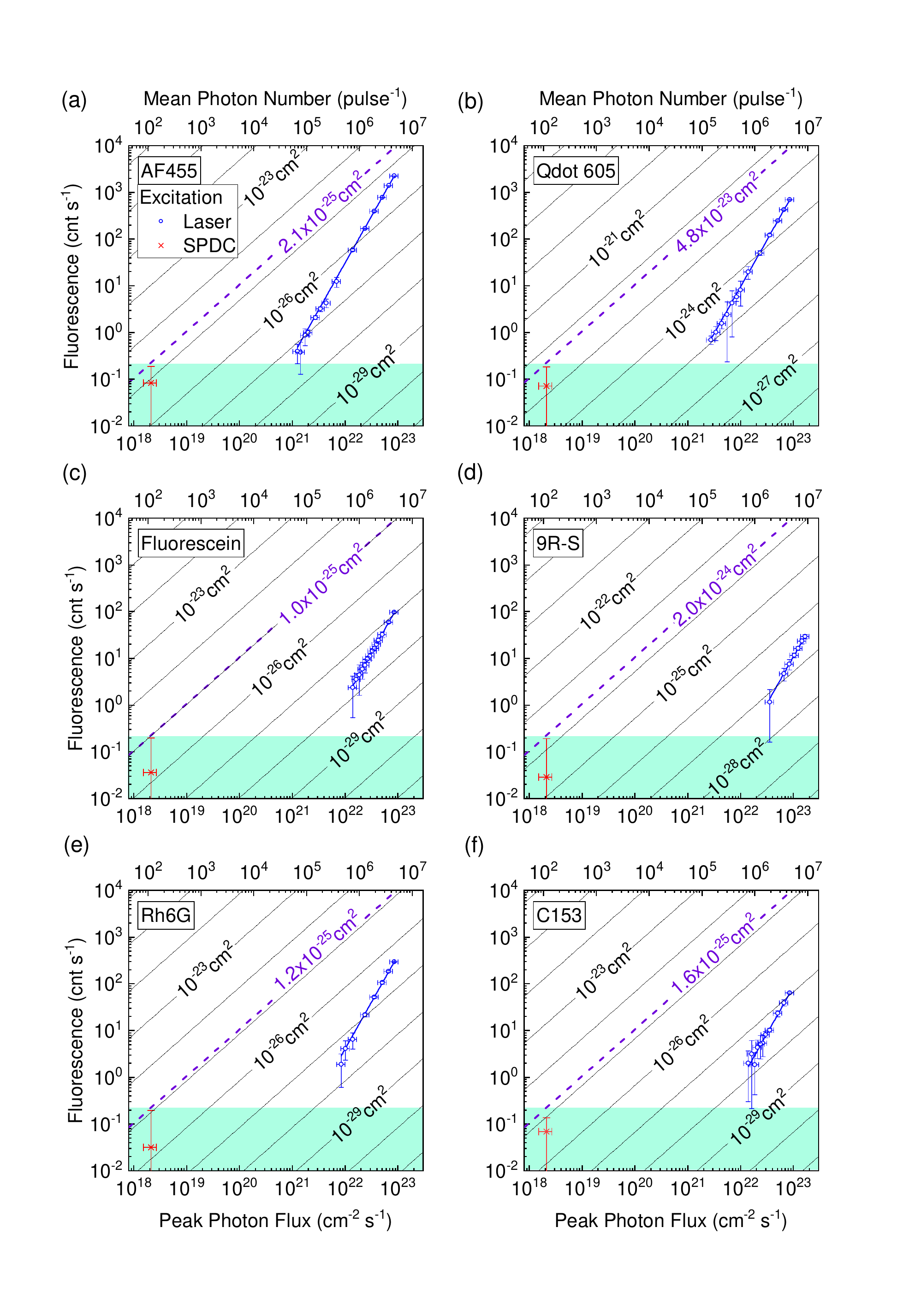}
\caption{\label{fig:sensPlot}Measured (blue data points), fit (blue solid lines) and calculated (black solid and purple dashed lines) fluorescence signal (left vertical axis in cnt~s$^{-1}$) for (a) $1.10\times10^{-3}$~mol~L$^{-1}$ AF455 in toluene, (b) $8\times10^{-6}$~mol~L$^{-1}$ qdot 605 in borate buffer, (c) $1.10\times10^{-3}$~mol~L$^{-1}$ fluorescein in pH 11 water, (d) $3.90\times10^{-4}$~mol~L$^{-1}$ 9R-S in chloroform, (e) $1.50\times10^{-3}$~mol~L$^{-1}$ Rh6G in methanol and (f) $1.10\times10^{-3}$~mol~L$^{-1}$ C153 in toluene. The bottom horizontal axis corresponds to the peak photon flux (photons~cm$^{-2}$~s$^{-1}$) of the coherent source (laser) (blue data points) or SPDC source (red data points). On the upper horizontal axis we show the SPDC mean photon number~\protect\footnotemark[1] (photons~pulse$^{-1}$), which corresponds to the peak photon flux on the lower horizontal axis. A signal below 0.22~cnt~s$^{-1}$ is indistinguishable from zero (green region). All E2PEF measurements produce a null result. Solid diagonal black lines show the calculated fluorescence count rate expected for various potential entangled two-photon absorption cross-sections in order-of-magnitude increments (cross-section noted along selected lines) for each fluorophore, assuming that the absorption rate is composed of only the linear photon-flux-dependent term. The purple dashed diagonal line represents the calculated signal using $\sigma_E^{\mathrm{UB}}$ (noted in purple) for each sample.~\protect\footnotemark[2]}
\footnotetext[1]{The conversion factor from mean photon number to peak photon flux is different for the coherent source because of its shorter pulse duration and smaller beam size. The laser conversion factor differs from that for SPDC by a factor of 16.7.}
\footnotetext[2]{We have omitted ``photons" in the units of peak photon flux and mean photon number, and ``fluorophore$^{-1}$" in the unit of the cross-section to improve the readability of the figure.}
\end{figure*}

We measure C2PEF over a range of photon fluxes for all six fluorophores. We use the fit to our experimental data to derive C2PA cross-sections (details in Appendix~\ref{DataAn}, values in Table~\ref{Tab:e2PA limits}). The values strongly agree with the cross-sections reported in literature. For all six fluorophores we are unable to discern an E2PEF signal. We use our measurable fluorescence lower bound to derive upper bounds on the E2PA cross-sections (details in Appendix~\ref{DataAn}, values in Table~\ref{Tab:e2PA limits}).

Figure~\ref{fig:sensPlot} shows measured fluorescence count rates as a function of peak photon flux for both laser (blue symbols) and SPDC (red symbols) excitation for all six fluorophores on log-log plots. For all samples, we find the fit (blue line) to the C2PEF signal to have a quadratic power dependence (with exponents in the range $1.95-2.05$); the signals are thus free of spurious events such as 1PEF or scattered light. For AF455 (Fig.~\ref{fig:sensPlot}(a)), we measure C2PEF down to the lowest peak photon flux of all the samples, $1.3\times10^{21}$~photons~cm$^{-2}$~s$^{-1}$, which is only 620 times larger than our SPDC peak photon flux. The C2PEF of fluorescein, 9R-S, Rh6G and C153 (Fig.~\ref{fig:sensPlot}(c)-(f)) is observed at a minimum flux approximately a factor of 10 higher than for AF455 and qdot 605 (Fig.~\ref{fig:sensPlot}(a),(b)). This minimum flux could be extended to lower values (but not as low as AF455 or qdot 605) if a longer integration time were used for the measurements. Fluorescence signals as low as 0.22~cnt~s$^{-1}$ should be measurable in our experiment (Appendix~\ref{DataAcq}). We denote this measurable fluorescence lower bound as $F^{\mathrm{LB}}$. A signal below this level is masked by the noise floor. As mentioned above, we do not observe E2PEF for any of the studied samples. This is demonstrated by the SPDC excitation data points (shown in red) lying below the noise floor (the green region in Fig.~\ref{fig:sensPlot}). For these measurements we use 30~mW pump power, which is just below the damage threshold of the SPDC crystal, to generate an SPDC peak photon flux of $2.1\times10^{18}$~photons~cm$^{-2}$~s$^{-1}$. 

We use our experimental characterizations and the component of the E2PA rate that depends linearly on excitation flux to calculate E2PEF signals for various potential values of $\sigma_{E}$ (Appendix~\ref{DataAn}). The results of these calculations are displayed as black diagonal lines in Fig.~\ref{fig:sensPlot}, with the corresponding $\sigma_{E}$ value noted along selected lines. The purple dashed diagonal line corresponds to the fluorescence signal calculated using the cross-section that produces $F^{\mathrm{LB}}$ at the peak photon flux of our SPDC source. We denote this cross-section the E2PA cross-section upper bound, $\sigma^{\mathrm{UB}}_{E}$. A summary of $\sigma^{\mathrm{UB}}_{E}$ values is given in Table~\ref{Tab:e2PA limits} and written in purple along the dashed diagonal lines. The sample fluorescein has the lowest $\sigma^{\mathrm{UB}}_{E}$ of $1.0\pm0.2\times10^{-25}$~cm$^2$~fluorophore$^{-1}$. The values of $\sigma^{\mathrm{UB}}_{E}$ for Rh6G, C153 and AF455 differ by less than or nearly a factor of two from that for fluorescein. Many of the parameters for these four samples are similar in magnitude: concentration, quantum yield, and the overlap of the emission spectra with the fluorescence collection system's transmittance spectrum (details on these parameters in Appendix~\ref{DataAn}). For 9R-S, the upper bound is one order of magnitude larger, which results from the poor overlap of the emission and system transmittance spectra. For qdot 605, the upper bound is a factor of 24 larger than for 9R-S. This is a result of poor spectral overlap, in addition to a sample concentration two orders of magnitude lower than that used for all other samples. As recommended by the supplier, we use the concentration of qdot 605 as received to avoid compromising the chemical stability of the sample.

The upper bounds our measurements place on $\sigma_E$ range from $10^{-25}$ to $\approx 5\times10^{-23}$~cm$^2$~fluorophore$^{-1}$. These are in stark contrast to the previously reported $\sigma_E$ values of $10^{-21}-10^{-16}$~cm$^2$~fluorophore$^{-1}$ shown in Table~\ref{Tab:e2PALiterature}. A particularly illuminating comparison can be made between our result and the published result for samples 9R-S and Rh6G. Using the previously reported $\sigma_E$ values, we estimate the expected E2PEF count rate in our setup. Assuming sample 9R-S has $\sigma_E \approx 2.4\times10^{-19}$~cm$^2$~fluorophore$^{-1}$~\cite{Eshun}, our calculations predict an E2PEF signal of $2.6\times10^4$~cnt~s$^{-1}$. For Rh6G, a value of $\sigma_E \approx 1.5\times10^{-21}$~cm$^2$~fluorophore$^{-1}$~\cite{Tabakaev} predicts an E2PEF signal of $2.7\times10^3$~cnt~s$^{-1}$~\footnote{Ref.~\cite{Tabakaev} used 1064~nm excitation whereas we excite at 810~nm. If $\sigma_{E}$ follows the same dependence on excitation wavelength as $\sigma_C$, 810~nm excitation should be more efficient by a factor of 7~\cite{Makarov2008}}. In either case, we actually measure a signal that is indistinguishable from zero, which is at least three to five orders of magnitude smaller than expected based on prior reports. We are able to reach such a high cross-section sensitivity in part because of our relatively large incident SPDC photon rate ($8.9\times10^{9}$~photons~s$^{-1}$).

Although $A_e$ and $T_e$ likely vary between experiments, we have no reason to believe these parameters alone differ by the many orders of magnitude required to explain this discrepancy. However, because $A_e$ and $T_e$ alter $\sigma_E$ and because the role of these parameters is not completely understood for E2PA in molecules, $A_e$ and $T_e$ should be reported alongside $\sigma_E$ values whenever possible. Large oscillations in $\sigma_{E}$ as a function of $T_e$ (``entangled two-photon transparencies"~\cite{Fei1997}) have been theoretically predicted for some molecular fluorophores~\cite{Schatz2020,Kojima2004,Burdick2018}. The fluorophores in our study have not been investigated, however it seems improbable that we probe orders-of-magnitude deep cross-section minima for all six fluorophores. There are other experimental parameters that vary between experiments, such as pump laser and SPDC spectral and temporal widths and the SPDC crystal characteristics. The effects of these differences are not well known and need more thorough study.

Table~\ref{Tab:e2PA limits} also shows estimates of the E2PA cross-section for the six fluorophores, $\sigma^{\mathrm{est}}_{E}$. These estimates are based on the relation given in Eq.~\eqref{e2PAtoc2PA} using our derived $\sigma^{\mathrm{exp}}_{C}$ given in Table~\ref{Tab:e2PA limits} and our estimates of $T_e$ and $A_e$ specified in Section~\ref{Exp}. We use the lower bound of $A_e$ in this estimation to show the largest value $\sigma^{\mathrm{est}}_{E}$ could take on. Although we do not anticipate Eq.~\eqref{e2PAtoc2PA} to yield an exact result, this estimate can provide useful insight about $\sigma_E$ values in a similar manner to the estimates of ionization cross-sections for atoms in Ref.~\cite{Mainfary1984}. These $\sigma^{\mathrm{est}}_{E}$ values are three to five orders of magnitude below our established cross-section upper bounds. These estimates provide a reference for the cross-section sensitivity necessary to observe E2PA.

We use our C2PEF and E2PEF results to determine an upper bound on the ``quantum advantage" of 2PA (QA$^{\mathrm{UB}}$). As previously mentioned, we define the quantum advantage as the ratio of the minimum photon flux required to observe C2PA to that for E2PA. By extrapolating our C2PEF fit to $F^{\mathrm{LB}}$ for the sample AF455, for example, we determine that C2PEF should be measurable down to $8.5\times10^{20}$~photons~cm$^{-2}$~s$^{-1}$. E2PEF is not measurable at our maximum SPDC photon flux, $2.1\times10^{18}$~photons~cm$^{-2}$~s$^{-1}$, but might be measurable at a higher photon flux. Thus, QA$^{\mathrm{UB}}$ of 2PA for this sample is 410. Values of QA$^{\mathrm{UB}}$ for all the samples (Table~\ref{Tab:e2PA limits}) range from $410-7000$, in contrast with QA of nearly $10^{10}$ in previous reports~\cite{Harpham2009,Villabona-Monsalve2018}. It is worth mentioning that although the QA can be increased if $A_e$ and $T_e$ are decreased while all other excitation parameters are held fixed, a many-orders-of-magnitude increase is unlikely.  

There are other publications in this field that support our findings. In particular, Ashkenazy \textit{et al.}~\cite{Ashkenazy2019} argued that using ``typical" values of $A_e$ ($50\,\mathrm{\mu m}^2$) and $T_e$ (50~fs), they can estimate $\sigma_{E} \approx 10^{-29} $~cm$^2$~fluorophore$^{-1}$ for metallic nanoparticles with a large C2PA cross-section ($\sigma_{C} \approx 100 $~GM at 1050 nm). Cross-sections of this size are in agreement with our established bounds of $\sigma_{E}$. Another interesting example is provided in the recent work by Li \textit{et al.}~\cite{Li2020} who used a single setup to measure both C2PEF and squeezed-light 2PEF (SL2PEF) of the samples DCM in dimethyl sulfoxide and fluorescein in pH 13 water. The squeezed light generated by four-wave mixing in a Rubidium vapor cell was varied over the range of $10^{13}-10^{16}$~photons~s$^{-1}$ (compare to our $\approx$~10$^{10}$~SPDC~photons~s$^{-1}$). The SL2PEF signals from DCM and fluorescein are factors of $\approx2.0-2.8$ and $\approx47$ larger, respectively, than the C2PEF signals at the same excitation flux. The authors did not report values for cross-sections. However, these significant but modest enhancements and the fact that measurements were performed with a squeezed light source that provides orders-of-magnitude higher photon rate than an SPDC source, are consistent with the upper bounds established in our study.

\section{Conclusions}
In this report, we discussed important aspects of designing and implementing a fluorescence-based E2PA measurement. We presented an experimental apparatus for measuring E2PEF and C2PEF in nearly identical experimental conditions. The results from C2PEF serve as a vital reference point for the capability of our fluorescence system. Although we do not observe an E2PEF signal, our results set upper bounds on $\sigma_{E}$ of the six chosen fluorophores in the range of $10^{-25}-5\times10^{-23}$~cm$^2$~fluorophore$^{-1}$. Two of these samples have published $\sigma_{E}$ values that are four and five orders of magnitude larger than the upper bounds we report. 

We emphasize that $\sigma_{E}$ depends on spatio-temporal properties of the excitation source, unlike $\sigma_{C}$. Without knowing the entanglement area and entanglement time, there is significant ambiguity in comparing cross-sections measured in different experimental apparatuses. For our source, we estimated a range within which our entanglement area is constrained, $2.1-13,700\,\mu$m$^2$, and we estimated the entanglement time, 1620~fs, based on our measured SPDC spectrum and estimated group delay dispersion. While we had hoped to measure these quantities directly, in lieu of this we made explicit the details of our setup and the assumptions that went into the estimation of these quantities.

Our results differ significantly from previous E2PA publications using SPDC excitation. Our evidence indicates that E2PA cross-sections are orders of magnitude smaller than previously claimed~\cite{Lee2006,Upton,Varnavski,Harpham2009,Guzman2010,Villabona-Monsalve2018,Eshun,Varnavski2020,Villabona-Monsalve2020,Villabona-Monsalve2017,Tabakaev}. As we demonstrated in this report, the clarification of the inconsistencies in the field is underway. This is an important step forward in the quantification of the achievable ``quantum advantage" and thus the merit of E2PA for spectroscopy and imaging applications.
\section*{Acknowledgments}
This work was supported by NIST and by the NSF Physics Frontier Center at JILA (PHY 1734006) and by the NSF-STROBE center (DMR 1548924). We thank D. Plusquellic, J. Squier, S. Polyakov, S. W. Nam, A. Migdall, A. Stefanov, J. Sipe, M. Ritsch-Marte  for valuable suggestions about designing the experiment and informative discussions about the E2PA process. We are grateful to T. Loon-Seng Tan and T. Cooper as well as T. Goodson for providing molecular samples. We also thank R. Thew, T. Goodson, A. Valencia and their group members for discussing their E2PA experiments with us. Certain commercial equipment, instruments, or materials are identified in this paper in order to specify the experimental procedure adequately. Such identification is not intended to imply recommendation or endorsement by NIST, nor is it intended to imply that the materials or equipment identified are necessarily the best available for the purpose.

K.M.P and A.M. contributed equally to this work.
\appendix
\begin{figure*}[tb]
\includegraphics[trim=0 10 0 0,clip,width=0.90\textwidth]{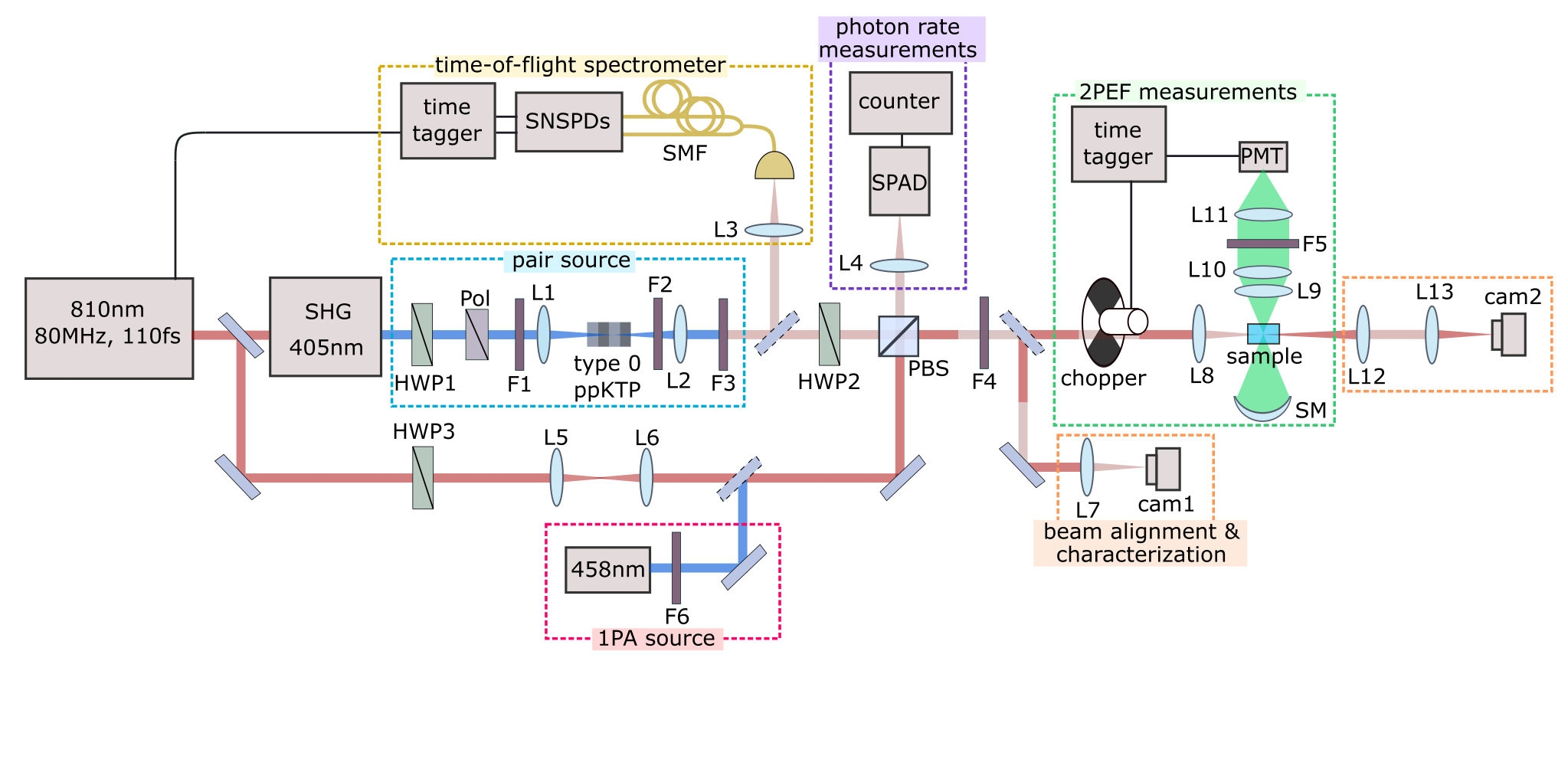}
\caption{\label{fig:setup_detailed}Detailed diagram of our experimental setup. See main text for abbreviation definitions and part numbers.}
\end{figure*}

\section{Detailed experimental setup and parts}
\label{ExpDetails}
In Fig.~\ref{fig:setup_detailed}, we show a detailed diagram of our setup with labeled parts. We list the part numbers below.\\

\noindent Main source
\setlist[enumerate]{itemsep=0mm}
\begin{enumerate}[label=--]
\item Laser source = Coherent Chameleon Discovery
\item SHG = APE HarmoniXX SHG
\end{enumerate}

\noindent Pair source
\begin{enumerate}[label=--]
\item HWP1 = zero-order half-wave plate 405~nm (Thorlabs WPH05M-405)
\item Pol = glan laser calcite polarizer (Thorlabs GL10-A)
\item F1 = dichroic mirrors (3 x 10Q20BB.1 and TLM-400-45S-1025), interference bandpass filters (2 x Thorlabs FBH405-10, 1 x Thorlabs FB405-10, 1 x Semrock FF01-405/10-25) and colored glass filter (Thorlabs FGB37M)
\item L1 = 300~mm focal length lens (Thorlabs LA4579-A)
\item ppKTP crystal (Raicol Crystals Ltd., type-0 SHG, AR coated, 3.425~$\mu$m poling period, 10~mm long)
\item crystal temperature controller (Covesion PV10) set to $30.00^{\circ}\mathrm{C} \pm 0.01^{\circ}\mathrm{C}$
\item F2 = interference longpass filters (Semrock BLP01-442R-25, BLP01-633R-25 and 3 x FF01-496/LP)
\item L2 = 200~mm focal length lens (Thorlabs LA1979-B) 
\item F3 = dichroic mirrors (2 x ARO MR6040) and interference longpass filter (Thorlabs FELH0700)
\end{enumerate}

\noindent Time-of-flight spectrometer (details on use in Appendix~\ref{TimeAndArea})
\begin{enumerate}[label=--]
\item L3 = 12.7~mm focal length achromatic doublet (Thorlabs AC064-013-B)
\item fiber beamsplitter (Thorlabs FC830-5OB-FC)
\item SMF = 2 x 500~m-long single mode fiber (Nufern 780-OCT)
\item SNSPDs = superconducting nanowire single-photon detectors (Quantum Opus, LLC, Opus One, optimized for the 850-1200~nm wavelength region) with a detection efficiency of $\approx75\%$ at 810~nm
\item closed-cycle helium cryocooler (Sumitomo HC-4E2)
\item temperature monitor (SIM 922)
\item detector bias and readout modules (Quantum Opus, LLC, QO-SIM-CRYO) 
\item time tagger = picosecond event timer and time-correlated single photon counting system (PicoQuant HydraHarp 400)
\end{enumerate}

\noindent Photon rate measurements (details on use in Appendix~\ref{MeanPhotNum})
\begin{enumerate}[label=--]
\item L4 = 50.2~mm focal length lens (Newport KPX082AR.16)
\item SPAD = single-photon avalanche diode (PerkinElmer SPCM-AQR-14)
\item counter = timer/counter/anaylzer (Tektronix FCA3103)
\end{enumerate}

\noindent 2PEF measurements (details on procedures in Appendix~\ref{DataAcq})
\begin{enumerate}[label=--]
\item optical chopper head and controller (New Focus 3501 Optical Chopper) 
\item L8 = 50~mm focal length lens (Thorlabs LA1131-B)
\item UV quartz sample cuvette with 2~mm width $\times$ 10~mm path length (FireFlySci, 1FLUV2), the narrow width is chosen to reduce fluorescence self-absorption in the sample
\item machined cuvette holder designed for stability and low footprint to bring optics close to excitation volume
\item L9, L10, L11 = Collection Optic with High Numerical Aperture (COHNA) lens system~\cite{Squier}
\item F5 = shortpass filter (Semrock FF01-758/SP-25) and sample-dependent bandpass filter (AF455 and C153 - Semrock FF02-470/100-25, qdot 605 and 9R-S - Chroma ET610/75m, fluorescein and Rh6G - Semrock FF01-535/150-25) (filter spectra is shown in Fig.~\ref{fig:spectralOverlap})
\item SM = spherical mirror with 15~mm focal length, 35~mm diameter (Edmund Optics, $\#$43-467)
\item PMT = photon-counting metal package photomultiplier tube (Hamamatsu H10682-210) 
\item thermoelectric cooler (TEC) (CP40336) to cool PMT to 5$^{\circ}$C 
\item CPU cooler (Rosewell PB120) for heat sink of TEC
\item time tagger = picosecond event timer and time-correlated single photon counting system (PicoQuant HydraHarp 400)
\end{enumerate}

\noindent Beam alignment and characterization (details on use in Appendix~\ref{Align})
\begin{enumerate}[label=--]
\item L7 = 50~mm focal length lens (Thorlabs LA1131-B)
\item L12 = 50~mm focal length lens (Thorlabs LA1131-B) 
\item L13 = 62.9~mm focal length lens (Newport KPX085AR.16)
\item cam1 = UI-3590LE-C-HQ camera
\item cam2 = Thorlabs UI-224XSE camera
\end{enumerate}

\noindent 1PA source (details on use in Appendix~\ref{CE})
\begin{enumerate}[label=--]
\item 458~nm source = OBIS 458 LX
\item F6 = neutral density (ND) filter wheel (Thorlabs)
\end{enumerate}

\noindent Other parts
\begin{enumerate}[label=--]
\item HWP2 = zero-order half-wave plate 808~nm (Thorlabs WPH10M-808)
\item HWP3 = half-wave plate 800~nm (Tower Optical)
\item L5 = 88.3~mm focal length lens (Newport KPX091AR.16)
\item L6 = 75~mm focal length lens (Newport KPC037AR.16)
\item PBS = polarizing beam splitting cube (Thorlabs PBS122) 
\item F4 = longpass interference filter (FELH0700)
\end{enumerate}

\section{Alignment details}
\label{Align}
A telescope (L5 and L6) is used to resize the laser beam close to the SPDC beam size at their foci in the sample. The alignment of the beam into the sample is checked using two cameras (cam1 and cam2). Lenses L7 and L8 are placed approximately the same distance from the flip mirror, enabling a view on cam1 of the beams at and near their foci in the cuvette. With this camera, we check the alignment of the beams through alignment irises, and measure beam size, Rayleigh range and overlap of the laser and SPDC beams. Two lenses after the sample (L12 and L13) collimate and focus either beam onto cam2. With this camera, we verify that the beams remain overlapped after passage through the cuvette, are centered along the $x$-direction inside of the cuvette and propagate nearly perpendicular to the cuvette walls they are incident on. To check that the beams are centered in the $x$-direction, we first use a translation stage to translate the cuvette along this axis and observe on the camera when the beams' strike the walls of the cuvette. We translate the micrometer to the midpoint of the locations of the wall striking events. To check that the beams propagate perpendicular to the walls they are incident on, we ensure that adding the cuvette does not displace the beams in the $x$ and $y$-directions significantly.

A typical transverse spatial overlap of the two beams at their foci (cam1) in the sample is shown in Fig.~\ref{fig:collectEff}b. The centers of the laser and SPDC beam are displaced from one another by $\approx5\,\mu$m vertically and horizontally. Zemax simulations (Fig.~\ref{fig:collectEff}) indicate that displacements of this magnitude have no effect on the collection efficiency. The beams' centers on cam2 are also overlapped within $\approx5\,\mu$m vertically and horizontally. The beam overlap is checked regularly. To initially align the beams in the $z$-direction, the lens L8 is placed roughly one focal length away from the center of the cuvette. 

The COHNA lens system (L9, L10 and L11) and filters (F5) are contained within a 25.4~mm-diameter lens tube. The spacing of the optics in the lens tube is based on Ref.~\cite{Squier}. The COHNA lens system (and filters) and the spherical mirror (SM) are each placed on a three-axis stage and initially aligned in the three directions based on the optimal spacings found using Zemax's OpticStudio. To optimize the alignment of the system (COHNA, SM and lens L8), we first adjust lens L8 to optimize collected C2PEF, which ensures that the excitation volume is centered with respect to the collection optics. Next, the COHNA lens system and the spherical mirror are each adjusted to maximize the collection of C2PEF. This process is iterated until the collection efficiency is optimal. A CW 458~nm source excites 1PEF in the sample to aid in the characterization of the geometrical collection efficiency (Appendix~\ref{CE}).

The alignment procedure using C2PEF optimizes the alignment of the system for C2PEF. For E2PEF, the alignment (only lens L8) must be slightly altered because we observe (on cam1) a shift between the foci of the laser and SPDC beam in the $z$-direction of $\approx500\,\mu$m. We compensate for this by shifting lens L8 so that either beam's focus is in the center of the cuvette prior to measurements.

\section{Sample preparation details}
\label{samplePrep}
The ``AF455" fluorophore~\cite{Kannan2004,Rogers2004} is provided by Drs. T. Loon-Seng Tan and T. Cooper from the Air Force Research Laboratory. Fluorescein, rhodamine 590 (6G) and coumarin 153 (540A) are ordered from Sigma-Aldrich and used as received. Qdot ITK Carboxyl Quantum dot 605 (qdot 605) in borate buffer is ordered from ThermoFisher, stored at 4$^{\circ}$C and only used for six months after receiving. The thienoacene fluorophore ``9R-S"~\cite{Eshun} is provided by Prof. T. Goodson from the University of Michigan. Various solvents are used to prepare the samples including toluene ($\ge99.98\%$), pH 11 water (Hydrion pH 11 buffer capsule in distilled water), methanol ($\ge99.9\%$), ethanol ($\ge99.5\%$), and chloroform ($\ge99.9\%$). The concentration and absorption/emission spectra of all samples (except qdot 605) are checked using a UV-VIS-NIR spectrophotometer (Agilent Cary 5000 Scan) and a fluorometer (Horiba Fluorolog-3 FL3-222). The absorption and emission spectra are compared with published spectra to ensure the samples are not contaminated or degraded. 

\section{Estimating entanglement time}
\label{TimeAndArea}
\begin{figure*}[t]
 \includegraphics[trim=30 10 5 5,clip,width=0.90\textwidth]{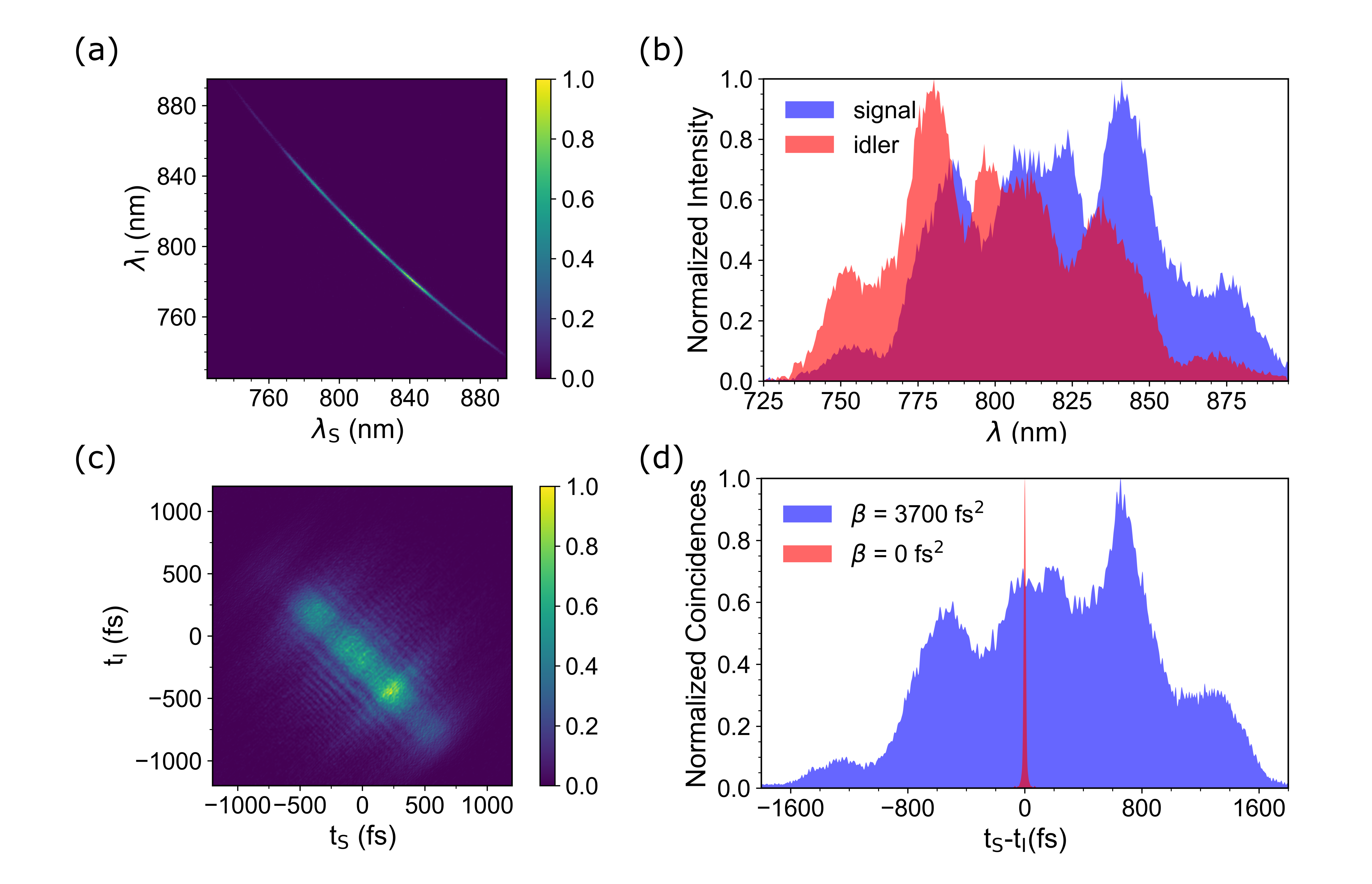}
 \caption{\label{fig:JointSpectrum} (a) Measured joint spectral intensity (JSI) where $\lambda_\mathrm{S,I}$ are the signal and idler wavelengths. (b) The JSI is projected onto the horizontal axis and vertical axis showing the signal (blue) and idler (red) spectra respectively. The FWHM of the signal and idler spectra are 79 and 72~nm respectively. The overlap of the spectra is evident in the dark red region. (c) Calculated joint temporal intensity (JTI) obtained through a discrete Fourier transform as described in the main text. (d) Projection onto the antidiagonal axis, t$_\mathrm{S}$-t$_\mathrm{I}$, of the JTI shown in (c) (blue) and for a transform-limited ($\beta =  0 \,\mathrm{fs}^2$) pulse (red). The FWHM of these projections are 1620~fs and 17~fs.}
 \end{figure*} 

The joint spectral intensity (JSI) distribution for our SPDC source is measured using a fiber-based time-of-flight spectrometer. We follow the procedure detailed in Refs.~\cite{Avenhaus2009,Gerrits2015}. We measure the dispersion of the fiber over the wavelength range from 680 to 1200~nm. At 810~nm, the center wavelength of the SPDC source, the fiber's dispersion is -0.114~ns~nm$^{-1}$~km$^{-1}$. Using this and the known timing information, we obtain the normalized JSI shown in Fig.~\ref{fig:JointSpectrum}(a). The shape indicates the expected wavelength anticorrelation of SPDC. Taking the projection of the JSI on the vertical and horizontal axes (for type-0 SPDC these can be called signal and idler projections, or vice versa) reveals several results, see Fig.\ref{fig:JointSpectrum}(b). The vertical and horizontal projections are shown in red and blue, respectively, with FWHM of 72 and 79~nm. The various features in the spectra and the detuning from degeneracy is likely a result of a combination of measurement artifacts, such as the spectral profiles of the optics.

The entanglement time $T_e$ is estimated as the FWHM of the antidiagonal projection of the joint temporal intensity (JTI) distribution~\cite{Fei1997}. We do not measure the JTI directly; instead we calculate it based on our measured JSI. Computing the JTI from the JSI requires knowledge of the spectral phase of the SPDC. We do not have a measurement of this phase; instead we estimate the accumulated group delay dispersion ($\beta$) of the pulse from the center of the crystal to the center of the sample to be 3700~fs$^2$. Our most dispersive elements are the polarizing beam splitter and the ppKTP crystal. We set the joint spectral amplitude (JSA) in the frequency domain to the square root of the JSI in the frequency domain multiplied by the phase factor due to $\beta$, $\mathrm{JSA} = \sqrt{\mathrm{JSI}} e^{i\beta(\omega_\mathrm{S}-\omega_\mathrm{P}/2)^2/2}e^{i\beta(\omega_\mathrm{I}-\omega_\mathrm{P}/2)^2/2}$ where $\omega_\mathrm{S}$, $\omega_\mathrm{I}$, and $\omega_\mathrm{P}$ are the frequencies of the signal, idler, and pump fields, respectively. In asserting this, we assume that the SPDC is transform limited in the center of the crystal and that the only significant accumulated phase factor is that due to $\beta$. We note that $\beta$ for signal photons would be distinct from that for idler photons if the two were instead orthogonally polarized~\cite{Valencia2002}. This is due to birefringence of various optical elements. We perform a discrete Fourier transform on the JSA to obtain the joint temporal amplitude (JTA). The magnitude squared of the JTA gives the JTI shown in Fig.~\ref{fig:JointSpectrum}(c). The projection of the JTI onto the antidiagonal (t$_\mathrm{S}$-t$_\mathrm{I}$) is shown in blue in Fig.~\ref{fig:JointSpectrum}(d). We find that $T_e \approx 1620\,\textrm{fs}$. This can be compared with a transform-limited ($\beta = 0\,\mathrm{fs}^2$) pulse (Fig.~\ref{fig:JointSpectrum}(d) in red) that has $T_e \approx 17\,\textrm{fs}$. The projections of the JTI onto horizontal and vertical axes both have FWHM's of 1040~fs. This width is a good approximation for the pulse duration of signal and idler beams because it is significantly larger than the pump pulse duration.

\begin{figure*}[t]
 \includegraphics[width=0.90\textwidth]{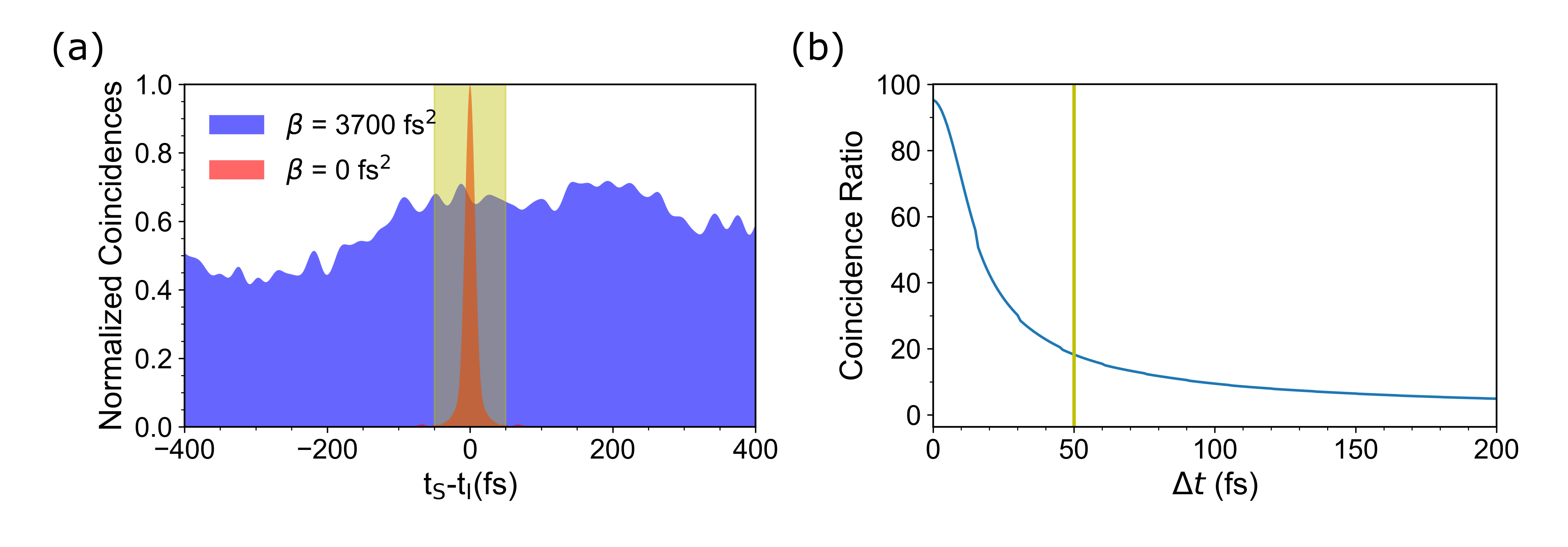}
 \caption{\label{fig:CoinRatio} (a) The JTI from Fig.~\ref{fig:JointSpectrum}(c) with a yellow filled-in region indicating a 50~fs time window (see main text for explanation). (b) The ratio of the number of coincidences of the transform-limited SPDC satisfying the relation $\left|\mathrm{t}_{\mathrm{S}}-\mathrm{t}_\mathrm{I}\right|\le \Delta \mathrm{t}$ to that for the dispersed SPDC ($\beta = 3700$~fs$^2$) as a function of $\Delta \mathrm{t}$. The yellow bar indicates the 50~fs time window shown in (a).}
\end{figure*} 

To make a probabilistic estimate of the advantage of dispersion compensation or a dispersion-free setup on the rate of E2PA in our experiment, we consider how dispersion affects the SPDC's fulfillment of the simultaneity requirement of 2PA. This simultaneity requirement asserts that two photons must arrive at the fluorophore within a time window ($\Delta \mathrm{t}$) set by the fluorophore's virtual state lifetime. For a dispersed pulse, fewer of the SPDC photon pairs arrive within this time window. A precise calculation of the virtual state lifetime of these large molecular fluorophores is not feasible, however we can consider any arbitrary time window. In Fig.~\ref{fig:CoinRatio}(a) the projection of the JTI from Fig.~\ref{fig:JointSpectrum}(d) is shown with a reduced range along the horizontal axis, along with a time window $\Delta \mathrm{t}=50\,\mathrm{fs}$ indicated in yellow. The number of photon pairs of the transform limited SPDC ($\beta = 0\,\mathrm{fs}$) which satisfy $\left|\mathrm{t}_{\mathrm{S}}-\mathrm{t}_\mathrm{I}\right| \le \Delta \mathrm{t}$ divided by the number of photon pairs of the dispersed SPDC ($\beta = 3700\,\mathrm{fs}$) which satisfy the same constraint (we call this the coincidence ratio) is shown in Fig.~\ref{fig:CoinRatio}(b) as a function of $\Delta \mathrm{t}$. A yellow bar indicates the 50~fs time window shown in Fig.~\ref{fig:CoinRatio}(a). For the smallest possible $\Delta \mathrm{t}$ we can consider based on our resolution (1~fs), the coincidence ratio is 95. Thus a factor of 95 more photon pairs of the transform-limited SPDC satisfy  $\left|\mathrm{t}_{\mathrm{S}}-\mathrm{t}_\mathrm{I}\right| \le 1$~fs than for the dispersed SPDC. This implies that for a virtual state lifetime of 1~fs, dispersion compensation or a dispersion-free setup would at most improve our E2PA rate by a factor of 95. If the virtual state lifetime is longer, the factor is smaller as indicated in Fig.~\ref{fig:CoinRatio}(b). This probabilistic analysis of the entanglement time's effect on the rate of E2PA ignores quantum interference effects predicted by more sophisticated theoretical models~\cite{Fei1997,Kojima2004,Burdick2018,Schatz2020}. The work on these models is ongoing and needs more thorough study. 

It is worth noting that we could reduce the $T_e$ of our source by reducing the SPDC bandwidth with a bandpass filter that is narrower than the $\approx$ 76~nm width. This would consequently reduce our photon flux. We do not try this as it seems unlikely that a decreased photon flux will increase our likelihood of measuring E2PA. 

\section{Effects of loss}
\label{loss}
In the case of linear loss between the SPDC generation crystal and the sample, the 2PA rate (Eq.~\eqref{Rg2}) is modified~\cite{Stevens2013}
\begin{equation}
     R = \kappa_2 \mathcal{T}^2 \left\langle \left.\hat{a}^\dagger\right.^2\hat{a}^2\right\rangle = \kappa_2 \mathcal{T}^2 \mu^2 g^{(2)},
\end{equation}
where the linear loss has been modeled as a lossless beamsplitter with transmittance $\mathcal{T}$. For excitation with a pulsed single-mode SPDC source,
\begin{equation}
R =  \frac{1}{2}\sigma_C \mathcal{T}^2 \left(\frac{\phi_{xtal}}{T A} + 3\phi_{xtal}^2 \right),
\end{equation}
where $\phi_{xtal}$ is the photon flux in the SPDC crystal (in our notation this is photon flux, not photon pair flux). Rewriting in terms of the photon flux at the sample ($\phi_{sample} = \mathcal{T} \phi_{xtal}$) yields
\begin{equation}
R_E =  \frac{1}{2}\left(\sigma_E \mathcal{T} \phi_{sample} +3 \sigma_C \phi_{sample}^2\right).
\end{equation}
To extract $\sigma_E$, the flux at the sample should be scaled by $\mathcal{T}$. If the flux is adjusted by attenuating the pump power, the first term in $R_E$ should scale linearly with $\phi_{sample}$. If the flux is instead adjusted by attenuating the downconverted light, this term should scale quadratically with $\phi_{sample}$. This signature should be present in either transmittance- or fluorescence-based measurement schemes. This loss-scaling signature was demonstrated in upconversion of downconversion by Dayan \textit{et al.}~\cite{Dayan2005,Dayan2007}, and can be used as a method to distinguish E2PA from other linear loss processes.

\section{Characterizing mean photon number}
\label{MeanPhotNum}
To characterize the SPDC mean photon number at the sample ($\mu_{sample}$), we rotate the half-wave plate (HWP2) just before the polarizing beam splitter in Fig. \ref{fig:setup_detailed} to direct all the SPDC light to a silicon single-photon avalanche diode (SPAD). We measure the count rate at low photon flux, and correct for the SPAD dead time and efficiency, the photon statistics of the downconversion source, and the difference in optical losses between the two paths. We perform this procedure for three low photon fluxes (where the SPAD is not saturated), and then extrapolate to the high flux used in the fluorescence measurements.

For a measured count rate $Q_{meas}$ on the SPAD, the measured click probability per laser pulse is 
\begin{equation}
    P_{click}^{meas} = \frac{Q_{meas}}{g},
\end{equation}
where $g = 8 \times 10^7$ pulses~s$^{-1}$ is the pulse repetition rate. Assuming a non-paralyzing dead time, the dead-time-corrected click probability can be found using~\cite{Neri2010}
\begin{equation}
    P_{click}^{corr} = \frac{P_{click}^{meas}}{1-N_{dead}P_{click}^{meas}},
\end{equation}
where $N_{dead}$ is the number of laser pulses the SPAD is dead for following detection of a photon. We measure the dead time of this SPAD as $\approx52$~ns, implying that $N_{dead}=4$. For a pump power of 50~$\mu$W, we measure $Q_{meas}=4.4\times10^6$~cnt~s$^{-1}$, and hence $P_{click}^{meas}=0.055$. The dead-time-corrected click probability is $P_{click}^{corr}=0.071$. This is the per-pulse click probability we would have expected to measure in the absence of dead time.

Converting this click probability to mean photon number requires knowledge of the system detection efficiency ($\eta_\mathrm{SDE}$) and the photon statistics of the SPDC source. These can be related through the expression~\cite{Kok2007}
\begin{equation}
\label{Pclick}
    P_{click}^{corr} = \sum_{n=1}^{\infty} [1 - (1 - \eta_\mathrm{SDE})^n] P(n),
\end{equation}
where $P(n)$ represents the probability that a pulse of SPDC light contains $n$ photons at the output of the downconversion crystal. At a wavelength of 810~nm, we calculate 83$\%$ cumulative transmittance of all the optics between the SPDC crystal and the SPAD based on manufacturer specifications. The manufacturer-specified efficiency of the SPAD is 55$\%$. Thus the system detection efficiency is $\eta_\mathrm{SDE} \approx 0.46$. The photon number distribution of a SMSV can be written~\cite{Gilles&Knight}
\begin{equation}
\label{PnSMSV}
    P(n) = \frac{\mu_{xtal}^{n/2} n!}{2^n (\frac{n}{2}!)^2(1+\mu_{xtal})^{(n+1)/2}}
\end{equation}
for $n$ even and $P(n) = 0$ for $n$ odd. Here $\mu_{xtal}$ is the mean photon number generated at the downconversion crystal. At 50~$\mu$W pump power ($P^{corr}_{click}=0.071$ and $\eta_\mathrm{SDE}=0.46$), we solve Eqs.~\eqref{Pclick} and \eqref{PnSMSV} for $\mu_{xtal}$. These numbers are consistent with a SMSV with mean photon number $\mu_{xtal}=0.22$.

\begin{figure*}[t]
\includegraphics[trim=35 0 35 80,clip,width=0.90\textwidth]{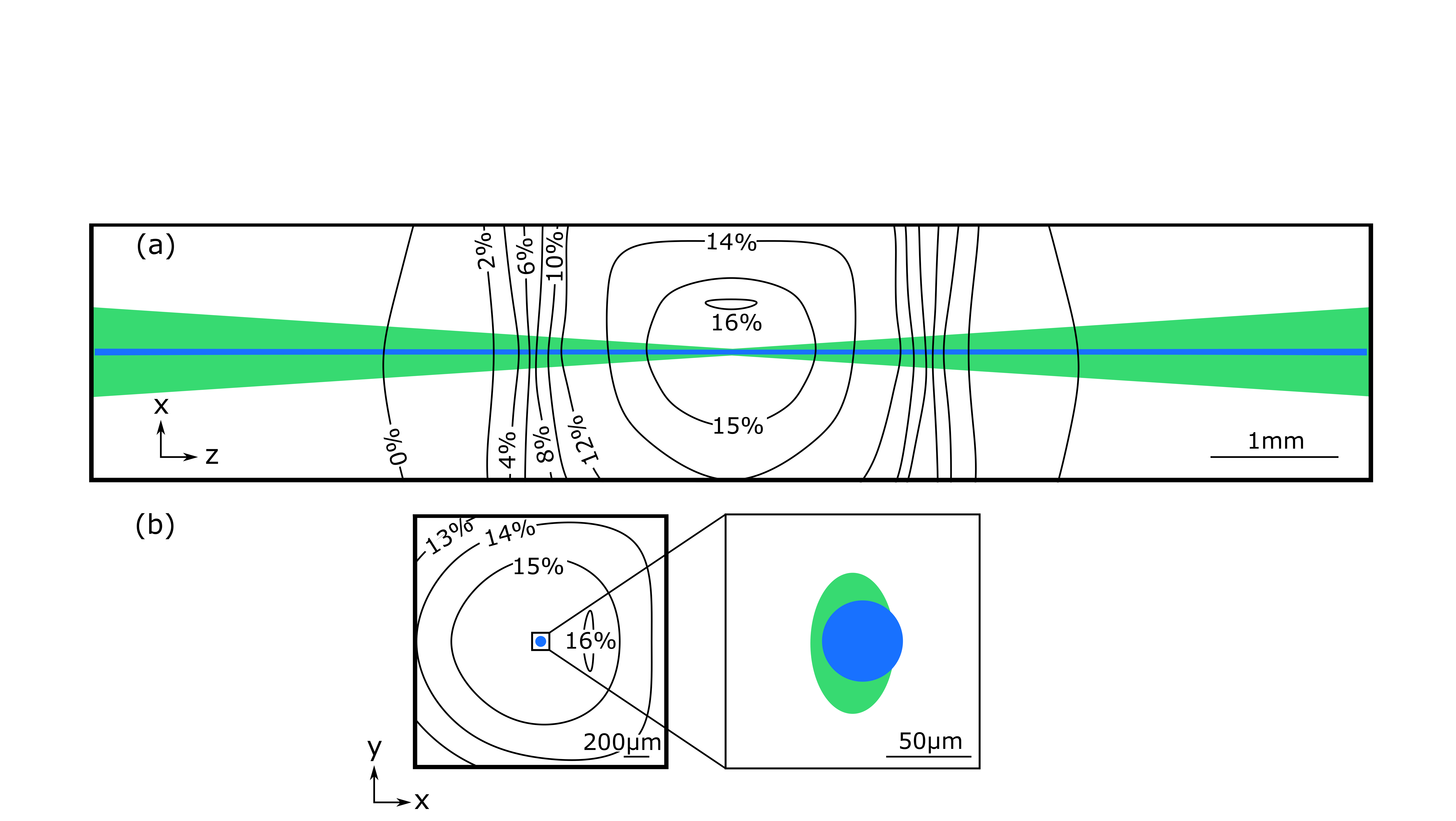}
\caption{\label{fig:collectEff}Illustration of the geometrical collection efficiency inside of the cuvette and the laser/SPDC beam overlap. (a) A cross-section of the cuvette in the $xz$ plane. The selected contours show where the collection efficiency is constant, based on Zemax simulations. The magnitude of the collection efficiency is scaled based on 1PEF measurements. The beam propagation is shown for the laser beam (blue) and the SPDC beam (green). (b) A cross-section of the cuvette in the $xy$ plane. The magnified image on the right is a view of the center of the $xy$ plane and shows approximate beam FWHMs and overlap.}
\end{figure*}

Although a SMSV can be a good first approximation to an SPDC source, our source emits into many spatial and spectral modes, as evidenced in part by the joint spectral measurements in Fig.~\ref{fig:JointSpectrum}(a). To approximate the many modes in our SPDC light, we perform a Bernoulli sampling of $M$ equally populated SMSVs, each with mean photon number $\mu_{xtal}/M$, assuming $\eta_\mathrm{SDE}=0.46$ for all modes. We substitute the resulting photon probability distribution $P(n)$ into Eq.~\eqref{Pclick} for a range of $M$ varying from 1 to 100, and find the resulting $\mu_{xtal}$ to only change in the range from 0.22 (one mode) to 0.21 (100 modes). Although we do not know exactly how many modes are present, the number of modes does not significantly impact the resulting value of $\mu_{xtal}$, so we use the value of 0.22 for 50~$\mu$W pump power.

We repeat this procedure at two other low pump powers (75 and 100~$\mu$W), and extrapolate the resulting linear fit to find $\mu_{xtal}\approx 147$~photons~pulse$^{-1}$ at the maximum pump power of 30~mW. Correcting this for the 24$\%$ loss between the SPDC crystal and the sample yields $\mu_{sample}\approx112$~photons~pulse$^{-1}$. This mean photon number is used to calculate the peak photon flux at the sample using Eq.~\eqref{peakflux}.

We can estimate the number of SPDC spectral modes based on the ratio of the SPDC to laser spectral widths, which gives $\approx8$ spectral modes. It follows that there are $\approx14$~photons~mode$^{-1}$~pulse$^{-1}$ at the sample. For these operating conditions, at which many pairs are spectrally and temporally overlapped, we may expect a significant contribution from C2PA to any measured signal.

\section{Fluorescence collection efficiency}
\label{CE}
Initial characterization of the fluorescence system's geometrical collection efficiency is performed using Zemax's OpticStudio. The solvent, glass cuvette walls, four collection optics and detector surface are modelled in the program. Using a merit function and an optimization algorithm, we find the ideal spacing of the optics. 

A 2PA process can only occur if two photons are sufficiently spatially overlapped at a fluorophore, thus the rate of C2PA and E2PA depend on the focusing of the respective beams. For C2PA, this is clearly evident through the quadratic photon flux dependence in the excitation rate (Eq.~\eqref{C2PArate}), where the photon flux depends inversely on the beam size. For E2PA, this spatial dependence is hidden, because the excitation rate depends linearly on photon flux (Eq.~\eqref{E2PArate}) in a similar manner to 1PA (a beam-size-independent process). The spatial dependence is instead included in the E2PA cross-section (Eq.~\eqref{e2PAtoc2PA}), which depends inversely on the entanglement area. 

In our experiment, the excitation beams are not collimated (see divergence of the SPDC (green) and laser (blue) beams in Fig.~\ref{fig:collectEff}(a)) and thus the excitation volume is a non-trivial shape. Furthermore the E2PA beam does not have a constant entanglement area (or E2PA cross-section) which complicates the ability to perform an exact calculation of the E2PA cross-section (or in our case an upper bound) for a given E2PA signal. To approximate this, we calculate the E2PA cross-section upper-bound in the region with nearly constant entanglement area (Appendix~\ref{DataAn}). For this calculation and our calculation of the C2PA cross-section (Appendix~\ref{DataAn}) it is critical to characterize the collection efficiency of our system as a function of the origin of the fluorescence along the $z$-direction. Ideally we would also take into account the collection efficiency's $x$- and $y$-dependence, however as we discuss below, this is less critical to the final result. 

In Zemax we simulate the collection efficiency as a function of the origin of the fluorescence within the cuvette volume. We model a point source of fluorescence that emits rays isotropically at some position in the cuvette. The number of those rays collected onto the detector are counted. We systematically translate this source in all directions to trace out contour plots of collection efficiency in the $xz$ plane (centered in $y$) (Fig.~\ref{fig:collectEff}(a)) and the $xy$ plane (centered in $z$) (Fig.~\ref{fig:collectEff}(b)). In this figure we rescale the collection efficiency found through Zemax to match experimental values, as discussed below. We find that the collection efficiency is slightly asymmetric in the $x$ direction, collecting slightly better when the point source is displaced towards the PMT. We ignore this minor asymmetry in the experiment and center the beams through the cuvette.

Although the spatial distributions of the excitation beams have some transverse extent, Fig.~\ref{fig:collectEff} shows that transverse displacements from $x=y=0$ must be large to significantly affect the collection efficiency ($>10^2\,\mu$m at $z=0$ and transverse displacements nearly negligible beyond $|z|>1$~mm). Our excitation beams' spatial distributions in the transverse directions are contained within a region of nearly constant collection efficiency, thus in our calculations (Appendix~\ref{DataAn}) we ignore the transverse spatial distribution of the excited fluorescence.

In Zemax we simulate the total collection efficiency ($\kappa$) of a particular excitation volume for the limiting cases of a uniform cylindrical excitation volume (50~$\mu$m diameter) that extends the length of the cuvette (centered in $x$ and $y$) and for that of a point source centered in the cuvette. For the former, the collection efficiency is at a minimum for the system (for a uniform excitation volume centered in the cuvette with 50~$\mu$m diameter), $\kappa_{\mathrm{min}} = 6.1\%$, and for the later, the collection efficiency is at a maximum for the system, $\kappa_{\mathrm{max}} = 20.2\%$

To a good approximation, the collection efficiency $K$ (found using Zemax) as a function of $z$ (cm) fits to a complementary error function. This can be qualitatively understood by the similarity of the simulation of the collection efficiency as a function of $z$ to a knife's edge beam profile measurement, which fits the same type of function. In both cases an intensity is measured as a function of the placement of an object. This object alters the intensity passed to a detector. Thus, the collection efficiency as a function of $z$ takes the form
\begin{equation}
\label{eq:CEeqn}
    K(z) = \frac{\kappa_{\mathrm{max}}}{2} \textrm{erfc} \left ( \alpha (|z|-z_0 ) \right ),
    \end{equation}
where $\kappa_{\mathrm{max}} = 0.20$, $\alpha = 2.8\,\mathrm{cm}^{-1}$ and $z_0 = 1.5\, \mathrm{cm}$. These parameters are set by Zemax collection efficiency simulations for the translation of a point source along the $z$ direction (centered in $x$ (cm) and $y$ (cm)), and $z = 0$ is the center of the cuvette. The function $K(z)$ is used to calculate the portions of an excitation volume extended along the $z$ direction that contribute to the collected fluorescence signal. Below we discuss our method to scale the collection efficiency as a function of z, $K(z)^{'}$ (where $^{'}$ indicates the experimental value rather than simulated), to fit the experimental conditions. 

We measure the minimum collection efficiency, $\kappa^{'}_{\mathrm{min}}$, of our fluorescence setup using a 1PEF-based technique. The results of this measurement are compared to the Zemax simulation of $\kappa_{\mathrm{min}}$ to scale the collection efficiency for experimental differences. As we mentioned, $\kappa^{'}_{\mathrm{min}}$ characterizes a system with a nearly uniform cylindrical excitation volume (50~$\mu$m diameter) that extends the length of the cuvette, thus we use an excitation source which generates an excitation volume of this kind.  

In a similar manner to the treatment in Ref.~\cite{Makarov2008}, the measured 1PEF rate, $F_{\mathrm{1}}$ (cnt~s$^{-1}$), can be described by
\begin{equation}
\label{1PEF}
    F_{\mathrm{1}} = N_1\frac{W}{h\nu}\kappa^{'}_{\mathrm{min}}\int\limits_{\lambda_i}^{\lambda_f}\gamma(\lambda)\Phi(\lambda)d\lambda,
\end{equation}
where $N_1$ (excitations~photon$^{-1}$) is the number of excitations per photon, $W$ (W) is the average power incident on the sample, $h\nu$ (J) is the average energy of an incident photon, $\lambda_{i,f}$ (nm) are initial and final wavelengths chosen to integrate over the entire emission spectrum of the sample, $\gamma(\lambda)$ is the wavelength-dependent component transmission efficiency (detector, filters, lenses and cuvette), and $\Phi(\lambda)$ (photons~excitation$^{-1}$~nm$^{-1}$) is the differential quantum yield. A normalization of quantum yield is used such that $\int_0^{\infty}\Phi(\lambda)d\lambda$ gives the value published in literature for the total quantum yield of the fluorophore (Table~\ref{Tab:sampleParam}). 

The number of excitations per photon is found using
\begin{equation}
\label{excitperphot}
   N_1 = 1-10^{-OD},
\end{equation}
where $OD= \epsilon  c  l$ is the optical density of the sample at the excitation wavelength, $\epsilon$ (L~mol$^{-1}$~cm$^{-1}$) is the extinction coefficient of the sample at the central excitation wavelength, $c$ (mol~L$^{-1}$) is the molar concentration of the sample and $l$ (cm) is the cuvette length.

We estimate $\gamma(\lambda)$,
\begin{eqnarray}
\label{eq:gamma}
    \gamma(\lambda) = \prod_{i=1}^N \mathcal{T}_{\textrm{filter}_i}(\lambda)  \prod_{j=1}^{M=3} \mathcal{T}_{\textrm{lens}_j}(\lambda)  \mathcal{T}_{\textrm{cuvette}}(\lambda) \nonumber \\ \times \frac{1}{2} \left ( 1+\mathcal{T}_{\textrm{cuvette}}^2(\lambda) \mathcal{R}_{\textrm{sph.mirror}}(\lambda) \right)  QE(\lambda),
\end{eqnarray}
where $\mathcal{T}(\lambda)$ and $\mathcal{R}(\lambda)$ are the transmittance and reflectance of a given optic and $QE(\lambda)$ is the PMT quantum efficiency. The various manufacturers' specifications are used to calculate $\gamma(\lambda)$. Here, we use one filter (bandpass) (F5) and thus $N=1$.  The laser, PMT quantum efficiency, fluorophore absorption and emission and filter transmittance spectra are shown in Fig~\ref{fig:spectralOverlap}(g)-(h).

To calculate $\kappa^{'}_{\mathrm{min}}$, we input Eq.~\eqref{excitperphot} into Eq.~\eqref{1PEF}, and solve for the minimum collection efficiency,
\begin{equation}
\label{minCE}
    \kappa^{'}_{\mathrm{min}} = \frac{F_1}{(1-10^{-OD}) W/(h\nu) \int\limits_{\lambda_i}^{\lambda_f}\gamma(\lambda)\Phi(\lambda)d\lambda}.
\end{equation}

The excitation source is a CW 458~nm laser. The beam FWHM and Rayleigh range at the focus on cam1 is measured to be 15~$\mu$m and 1~mm respectively. This Rayleigh range suggests that the beam size will be significantly larger at the edge of the 10~mm path length cuvette compared to at the center. However, the beam has a small transverse spatial extent for all $z$ (at $|z|=l/2$ the beam size is $\approx4\%$ of the cuvette width) relative to the collective efficiency contour spacing in the transverse direction. Thus, the excitation volume can be approximated as a uniform cylindrical volume that extends the length of the cuvette. Using a similar argument, although the beam size is smaller than that used in the simulation (50~$\mu$m), the difference can be neglected based on the relatively large spacing of the collection efficiency contours.

We use the samples Rh6G in ethanol and fluorescein in pH 11 water. The sample is prepared at a relatively low concentration ($\approx0.1-10\times10^{-6}$mol~L$^{-1}$) and the $OD$ is measured in a spectrophotometer. During measurements, the amount of power ($W$) reaching the sample is measured after L8 and varied using the ND wheel (F6) after the output of the laser. We first send the laser through the solvent to check that there is no signal due to scattered light. Next, the laser is sent through the sample and a signal is measured. The fluorescence signal is measured at six different excitation powers ranging from 10 - 150~nW. 

Using the comparison of the experimentally determined and simulated minimum collection efficiency, we rescale the maximum collection efficiency of the system, $\kappa^{'}_{\mathrm{max}}=\kappa^{'}_{\mathrm{min}}/\kappa_{\mathrm{min}}\times\kappa_{\mathrm{max}}$, which goes into the experimental $K(z)^{'}$ (same as Eq.~\eqref{eq:CEeqn}, but with $\kappa^{'}_{\mathrm{max}}$ instead of $\kappa_{\mathrm{max}}$). We measured an average $\kappa^{'}_{\mathrm{min}} = 3.9 \pm 0.6 \%$ and $5.4 \pm 0.7 \%$ for Rh6G and fluorescein respectively. Using the average of these two, we find $\kappa^{'}_{\mathrm{max}} = 15.4 \%$.
 
We note that in fluorescence measurements, especially those performed at high sample concentrations, fluorescence self-absorption can reduce the measured signal. In our experiment, our narrow cuvette width minimized this effect. Our measurements suggest that self-absorption is negligible.

\section{Data acquisition, error bars and measurable fluorescence lower bound}
\label{DataAcq}
In this section we describe the details of data acquisition for C2PEF and E2PEF measurements. First we describe our fluorescence background subtraction method. Next we describe how C2PEF measurements were performed. Afterwards we describe the choice of integration times for E2PEF measurements and how those measurements were performed. Lastly we describe how the measured quantities, error bars and measurable fluorescence lower bound on Fig.~\ref{fig:sensPlot} were determined.

The laser and SPDC beams are optically chopped to perform on-the-fly background subtraction on the fluorescence signal. The timetagger histogram, which shows counts registered on the PMT as a function of time, is separated into background and signal portions. The background portion (chopper blade blocking beam) is subtracted from the signal portion (chopper blade passing beam). We calibrate this background subtraction method using a strong C2PEF signal. For $\approx5\%$ of the measurement runtime the chopper blade is neither completely blocking nor passing the beam; this portion of the measurement is discarded. 

For C2PEF measurements, the laser power is controlled using a motorized half-wave plate (HWP3). The power is measured (Thorlabs S130C power sensor and PM100D meter) by flipping the sensor into the beam using a motorized flip mount that ensures repeatable positioning. The power sensor and meter are compared with a calibrated photodiode to determine the correction factor necessary for absolute power readings. At each power, $3-5$ C2PEF measurements are performed. The integration times at higher powers are 30 seconds and at lower powers are 30 minutes.

We characterize the stability of the fluorescence measurements using an Allan deviation analysis, and base our measurement integration time for the E2PEF measurements on the result. To do this, we place the $1.10\times10^{-3}$~mol~L$^{-1}$ fluorescein sample in the cuvette, unshutter the laser beam and measure the C2PEF signal every minute for one 14 hour period overnight and one 11 hour period during the day. We use this data to check the Allan deviation at various integration times. The Allan deviation is found to have a minimum at 45 minutes integration time. 

For E2PEF measurements, the SPDC pump laser power is set to 30~mW and monitored periodically. Three E2PEF measurements are performed on each sample. These measurements are each 45 minutes long. We also block the beam periodically and take a 45 minute background measurement. We compare these measurements with those with the beam unblocked to look for significant changes in the signal. We find no changes.

The C2PEF measurements are averaged for each sample at each power. The E2PEF measurements are averaged for each sample. These averages are displayed on Fig.~\ref{fig:sensPlot}. The corresponding vertical error bars are assigned in a systematic way. First, we compare the standard deviation of the set of measurements to the sets' uncertainty due to Poisson counting statistics. The larger of these two values is multiplied by two (coverage factor $k = 2$) and used for the vertical error bar. 

The horizontal error bars correspond to the uncertainty in peak photon flux (bottom axis), which is larger than the uncertainty in mean photon number (top axis). This larger uncertainty arises because of the additional uncertainty in the beam size and pulse duration. The uncertainty in the mean photon number, beam size and pulse duration is propagated to give an uncertainty in peak photon flux. A coverage factor $k = 2$ is again used to achieve $\approx 95\%$ confidence that the true value lies within the bounds set by the error bars. 

The measurable fluorescence lower bound, $F^{\mathrm{LB}}$, is assigned by first checking the results of C2PEF measurements at low photon flux and then by measuring ``zero signal" with SPDC excitation. We measure C2PEF at rates as low as $0.38\pm0.24$~cnt~s$^{-1}$ that agree well with the quadratic fit of the C2PEF data measured at higher excitation flux. This sets our confidence in signals at least as low as 0.38~cnt~s$^{-1}$. Next, we measure zero signal to determine what we should expect in the absence of signal. To do this, we place the $1.10\times10^{-3}$~mol~L$^{-1}$ fluorescein sample in the cuvette, unshutter the SPDC beam and subsequently block the SPDC beam using black aluminum foil tape (Thorlabs T205-1.0) placed after filters F3. We then acquire data for 405~minutes, or nine 45~minute measurements. The purpose of blocking the beam instead of shuttering it is to serve as an additional check for scattered light entering the detector. The fluorescein sample aids in this purpose by serving as a source that could be excited by either scattered or background light. The average of these measurements is $0.04^{+0.22}_{-0.04}$~cnt~s$^{-1}$. It was clear from these measurements that no stray signals enter the detector. From this, our $F^{\mathrm{LB}}$ is set to 0.22~cnt~s$^{-1}$ ($2\sigma$ from zero) with $\approx 95\%$ confidence. The value of $F^{\mathrm{LB}}$ sets the vertical position of the light green region in Fig.~\ref{fig:sensPlot}.

\section{Calculating upper bounds of the E2PA cross-section}
\label{DataAn}
Here we describe the equations relevant for the calculation of the E2PA cross-section upper bounds. First we describe the C2PEF signal and the derivation of C2PA cross-sections from the fit to our C2PEF data. Then we describe the E2PEF signal and the derivation of E2PA cross-section upper bounds based on our measurable fluorescence lower bound.

\begin{figure*}[p]
\includegraphics[trim=35 45 40 55,clip,width=0.65\textwidth]{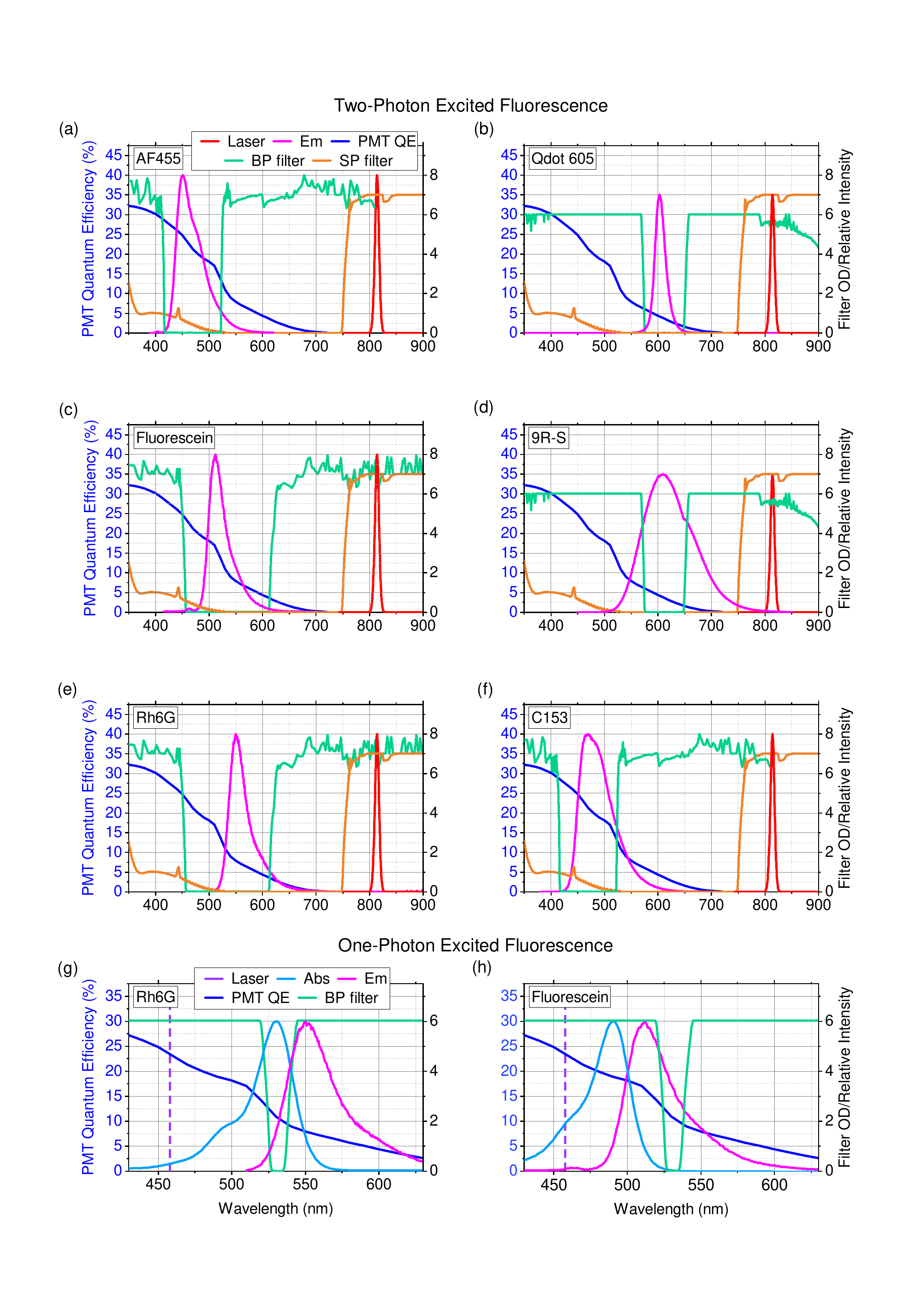}
\caption{\label{fig:spectralOverlap} Spectral overlap summary for two-photon excited fluorescence (2PEF) measurements (a)-(f) of samples (a) AF455 in toluene, (b) qdot 605 in borate buffer, (c) fluorescein in pH 11 water, (d) 9R-S in chloroform, (e) Rh6G in methanol and (f) C153 in toluene and one-photon excited fluorescence (1PEF) collection efficiency measurements (g)-(h) of samples (g) Rh6G in ethanol and (h) fluorescein in pH~11 water. For 2PEF, the laser (red), fluorophore emission (Em) (magenta), PMT quantum efficiency (QE) (blue), bandpass (BP) filter (light green) and shortpass (SP) filter (orange) spectra are shown. The laser spectrum is measured using a USB4000 OceanOptics spectrometer. The SPDC spectrum is shown in Fig.~\ref{fig:JointSpectrum}. For 1PEF, the laser (indigo), fluorophore absorption (Abs) (light blue) and emission (Em) (magenta), PMT QE (blue) and BP filter (light green) spectra are shown. The PMT QE is indicated along the left vertical axes, whereas all other spectra use the right vertical axes. For the filters, the right vertical axes indicate the filter optical density (OD), whereas for all other spectra the right vertical axes show a relative intensity. The relative intensities of the laser, absorption and emission are normalized to the height of the peak filter OD for the respective plot. The absorption and emission spectra are measured using a spectrophotometer and fluorometer, except for qdot 605 (data taken from ThermoFisher). The PMT QE is taken from Hamamatsu specifications. All filter spectra are from the manufacturer, except for the SP filter in the 350-550~nm range (spectrophotometer).}
\end{figure*}

The C2PEF signal, $F_C$ (cnt~s$^{-1}$), measured in our experiment can be described by
\begin{equation}
\label{c2PEF}
    F_C = g \int\limits_{-l/2}^{l/2} N_C(z)  K(z)^{'} dz \int\limits_{\lambda_i}^{\lambda_f}\gamma(\lambda)\Phi(\lambda) d\lambda,
\end{equation}
where $g$ (pulses~s$^{-1}$) is the pulse repetition rate, $l$ (cm) is the cuvette path length, $N_C(z)$ (excitations~cm$^{-1}$~pulse$^{-1}$) is the number of excitations per infinitesimal step $dz$ (cm) along the cuvette length per laser pulse, $K(z)^{'}$ is the geometrical collection efficiency as a function of $z$ (cm) as described in Eq.~\eqref{eq:CEeqn},  $\gamma(\lambda)$ is the component transmission efficiency as described in Eq.~\eqref{eq:gamma} (where here $N = 2$) and $\Phi(\lambda)$ (photon~excitation$^{-1}$~nm$^{-1}$) is the differential fluorescence quantum yield. A proper normalization of quantum yield is used such that $\Phi = \int_0^{\infty}\Phi(\lambda)d\lambda$ gives the value published in literature (Table~\ref{Tab:sampleParam}) for the total quantum yield of the fluorophore. The integration limits for the $\lambda$ (nm) integral are set so that the integral spans over the entire emission spectrum of the fluorophore. The laser, fluorophore emission, PMT quantum efficiency and filter spectra for 2PEF measurements are shown in Fig~\ref{fig:spectralOverlap}(a)-(f). 

The experimental conditions are such that ground state depletion and beam depletion are negligible~\cite{Rumi2010}, thus we define $N_C(z)$ as
\begin{equation}
\label{numberExcitations}
    N_C(z) = \frac{1}{2} \sigma_{C}  n \int\limits_{-1/2g}^{1/2g}\int\limits_{-\infty}^{\infty}\int\limits_{-\infty}^{\infty} \phi(x,y,z,t)^2 dx dy dt,
\end{equation}
where $\sigma_{C}$ (1 GM = 10$^{-50}$ cm$^4$~s~photon$^{-1}$~fluorophore$^{-1}$) is the C2PA cross-section, $n$ (fluorophores~cm$^{-3}$) is the number density of fluorophores and $\phi(x,y,z,t)$ (photons~cm$^{-2}$~s$^{-1}$) is the photon flux of the laser beam. The factor of $1/2$ carries units of excitations per photons absorbed. The $dx$ (cm) and $dy$ (cm) integrals extend over the entire beam and the $dt$ (fs) integral extends over the pulse repetition time ($1/g$). Equation~\eqref{numberExcitations}, is related to the familiar phenomenological C2PA excitation rate, $R$ (excitations~s$^{-1}$~fluorophore$^{-1}$), described in Eq.~\eqref{C2PArate} by
\begin{equation}
    N_C(z) = n \int\limits_{-1/2g}^{1/2g}\int\limits_{-\infty}^{\infty}\int\limits_{-\infty}^{\infty} R dx dy dt,
\end{equation}
with $R$ having implied dependence on $x$, $y$, $z$ and $t$.

The temporal and transverse spatial profiles of the laser beam or SPDC beam can be approximated by Gaussians. Assuming the laser is always on, $\phi(x,y,z,t)$ takes the form
\begin{eqnarray}
\label{flux}
\phi(x,y,z,t) = \phi_0(z)\nonumber \\\times\textrm{Exp}\left(-4\textrm{ln}2\left(\frac{x^2}{\Delta x(z)^2}+\frac{y^2}{\Delta y(z) ^2}\right)\right)\nonumber \\ \times \sum_{i=-\infty}^{\infty} \textrm{Exp}\left(-4\textrm{ln}2\frac{(t+i/g)^2}{\tau^2}\right) ,
\end{eqnarray}
where $\phi_0(z)$ (photons~cm$^{-2}$~s$^{-1}$) is the peak photon flux as a function of $z$, $\tau$ (fs) is the FWHM pulse duration and $\Delta x (z)$ (cm) and $\Delta y (z)$ (cm) are the FWHM beam widths. The photon flux has $z$ dependence because it is focused into the sample. The FWHM beam width in the $x$ direction, for example, varies as a function of $z$ as
\begin{equation}
\label{beamFWHM}
    \Delta x(z) = \Delta x_0 \sqrt{1+(z/z_R)^2},
\end{equation}
where $\Delta x_0$ (cm) is the beam FWHM at the focus and $z_R$ (cm) is the Rayleigh range.

We can define the average photon rate at the sample $Q$ (photons~s$^{-1}$) in terms of the photon flux $\phi(x,y,z,t)$,
\begin{equation}
\label{photonrate}
Q = g \int\limits_{-1/2g}^{1/2g}\int\limits_{-\infty}^{\infty}\int\limits_{-\infty}^{\infty}\phi(x,y,0,t) dx dy dt = \frac{W}{h\nu},
\end{equation}
where $W$ (W) is the average laser or SPDC power and $h\nu$ (J) is the average energy of an incident photon. Here we have arbitrarily chosen to use the photon flux at $z=0$. The peak photon flux, $\phi_0(z)$, can be found from Eq.~\eqref{photonrate} by performing the integration of $\phi(x,y,0,t)$ over $x$, $y$ and $t$,
\begin{eqnarray}
\label{peakflux}
    \phi_0(z) = \frac{W}{h\nu}\left(\frac{4\mathrm{ln}(2)}{\pi}\right)^{3/2}\frac{1}{\Delta x(z)\Delta y(z) g \tau} \nonumber \\ = \frac{2\sqrt{2} \mu}{T A(z)}.
\end{eqnarray}
The second equality emphasizes, in accordance with Section~\ref{theory}, that the peak photon flux can be expressed as $2\sqrt{2}$~\footnote{The factors of 2 and $\sqrt{2}$ scale the photon rate to the effective photon rate at the location of the beam's peak in space and time.} multiplied by the mean photon number at the sample $\mu = Q/g$ (photons~pulse$^{-1}$) divided by the effective mode area, $A(z)$ (cm$^2$), and the effective pulse duration, $T=\tau/\sqrt{2\mathrm{ln}(2)}$ (fs). The effective beam area as a function of $z$ is found through the $x$ and $y$ integration of the photon flux
\begin{equation}
    A(z) = \frac{\pi \Delta x (z) \Delta y (z)}{2 \mathrm{ln}(2)}.
\end{equation}

Using Eqs.~\eqref{numberExcitations},~\eqref{flux} and~\eqref{peakflux}, and performing the integration over x, y and t, we can rewrite Eq.~\eqref{c2PEF} in terms of the laser power
\begin{eqnarray}
\label{C2PEFsim}
    F_C = \sqrt{2} \left(\frac{\mathrm{ln}(2)}{\pi}\right)^{3/2} \frac{\sigma_{C}n W^2}{\tau g \left(h\nu\right)^2} \nonumber \\ \times\int\limits_{-l/2}^{l/2}\frac{K(z)^{'}}{\Delta x(z) \Delta y(z) }dz\int\limits_{\lambda_i}^{\lambda_f}\gamma(\lambda)\Phi(\lambda) d\lambda.
\end{eqnarray}
To derive the C2PA cross-section, we solve for $\sigma_C$ in Eq.~\eqref{C2PEFsim},
\begin{eqnarray}
\label{C2PAcrosssection}
    \sigma_C = \frac{1}{\sqrt{2}} \left(\frac{\pi}{\mathrm{ln}(2)}\right)^{3/2} \frac{\tau g \left(h\nu\right)^2}{n} \nonumber \\ \times \frac{F_C/W^2}{\int\limits_{-l/2}^{l/2}K(z)^{'}/(\Delta x(z) \Delta y(z)) dz\int\limits_{\lambda_i}^{\lambda_f}\gamma(\lambda)\Phi(\lambda) d\lambda}.
\end{eqnarray}

All the parameters in Eq.~\eqref{C2PAcrosssection} are known for our measurements through experiments, simulations and specifications. The parameter $\tau$ is measured using a SwampOptics Grenouille 8-50-USB, $g$ and $h\nu$ are specified by the laser manufacturer,  $n$ is measured (Appendix~\ref{samplePrep}), $K(z)^{'}$ is determined through Zemax and experimental verification (Appendix~\ref{CE}), $\Delta x_0$, $\Delta y_0$ and $z_R$ are measured (Appendix~\ref{Align}), $\gamma(\lambda)$ is calculated based on optics' specifications (Appendix~\ref{CE}), $\Phi(\lambda)$ (except in the case of AF455) is known from published measurements and $F_C/W^2$ (cnt~s$^{-1}$~$\mu$W$^{-2}$) is the fit to our experimental C2PEF data. Table~\ref{Tab:sampleParam} shows sample specific parameters and Table~\ref{Tab:expParam} shows apparatus parameters general for all samples. The results of our C2PEF measurements produce the experimental C2PA cross-sections (using Eq.~\eqref{C2PAcrosssection}), $\sigma_C^{\mathrm{exp}}$, shown in Table~\ref{Tab:e2PA limits}.
\begin{table}[ht] 
    \caption{\label{Tab:sampleParam}Summary of sample parameters}
    \begin{ruledtabular}
    \begin{tabular}{ ld{1}lc }
    Sample &  \multicolumn{1}{c}{$c \times 10^6$} & \multicolumn{1}{c}{$\Phi$~[Ref.]} & $\int\limits_{\lambda_i}^{\lambda_f}\gamma(\lambda)\Phi(\lambda) d\lambda/\Phi $ \\
    & \multicolumn{1}{c}{(mol~L$^{-1}$)} & \multicolumn{1}{c}{(photon~excitation$^{-1}$)} & \\ \hline
    AF455 & 1100 & 0.67~\cite{Rogers2004} & 0.0515 \\
    Qdot 605 & 8 & $0.74\pm0.04$~\cite{Gaigalas2014} & 0.0285 \\
    Fluorescein & 1100 & 0.93~\cite{Martin1975} & 0.0789  \\
    Rh6G & 1500 & 0.90~\cite{Penzkofer1987} & 0.0484 \\
    C153 & 1100 & $0.82\pm0.04$~\cite{Krolicki2002} & 0.0580 \\
    9R-S & 390 & 0.66~\cite{Eshun} & 0.0157 \\
    \end{tabular}
    \end{ruledtabular}
\end{table}

\begin{table}[ht]
    \caption{\label{Tab:expParam}Summary of apparatus parameters}
    \begin{ruledtabular}
    \begin{tabular}{ llP{2cm}P{2cm} }
    Parameter & \multicolumn{1}{c}{unit} & Laser & SPDC \\ \hline
    $\Delta x_0$ & $\mathrm{\mu m}$ &49 & 51  \\
    $\Delta y_0$ &$\mathrm{\mu m}$ &49 & 84 \\
    $z_R$ & mm &5.1 & 0.4  \\
    $\tau$ & fs &111 & 1040 \\
    $g$ & 10$^6$~pulses~s$^{-1}$ &  \multicolumn{2}{c}{80} \\
    $K(z)^{'}$ &  &  \multicolumn{2}{c}{$\frac{0.154}{2} \textrm{erfc} \left ( 2.78 (|z(\mathrm{c m})|-1.51 ) \right )$}  \\
    $Q$ & photons~s$^{-1}$& N/A & 8.9$\times10^9$\\
    $\mathcal{T}$ & & N/A & 0.76 \\
    $F^{\mathrm{LB}}$ & cnt~s$^{-1}$ & \multicolumn{2}{c}{0.22} \\
    \end{tabular}
    \end{ruledtabular}
\end{table}

If we assume the expected E2PEF signal depends only linearly on photon flux, we can estimate the E2PEF signal $F_E$ (cnt~s$^{-1}$) generated in our experiment as
\begin{equation}
\label{e2PEFrate}
    F_{E} = N_E g \int\limits_{-z_R}^{z_R}K(z)^{'} dz \int\limits_{\lambda_i}^{\lambda_f}\gamma(\lambda)\Phi(\lambda) d\lambda,
\end{equation}
where $N_E$ (excitations~cm$^{-1}$~pulse$^{-1}$) is the number of excitations per infinitesimal step $dz$ along the cuvette length per laser pulse. The parameter $N_E$ is defined as
\begin{equation}
\label{numberExcitationsE}
    N_E = \frac{1}{2} \sigma_{E} \mathcal{T} \frac{Q}{g} n ,
\end{equation}
where $\sigma_{E}$ (cm$^2$~fluorophore$^{-1}$) is the E2PA cross-section and $\mathcal{T}$ is the transmittance of the photons through all the optics between the center of the crystal and the center of the sample. The parameter $\mathcal{T}$ is included in $N_E$ but not $N_C$ because of the result found in Appendix~\ref{loss} (in this Appendix, $\phi$ and $Q$ are the values at the sample). As we mentioned in Section~\ref{theory} and Appendix~\ref{CE}, the dependence of the E2PA rate on the spatial overlap of photons is contained in the cross-section (unlike for C2PA) and thus a cross-section is only valid for a beam of constant entanglement area and thus size. Our SPDC beam is not collimated, instead we attempt to compensate for the changing entanglement area by setting the limits of the $z$ integral from $-z_R$ to $z_R$, which is the region that we expect the majority of a potential E2PEF signal to arise from and should have fairly uniform entanglement area and time.

We can rewrite Eq.~\eqref{e2PEFrate} using Eq.~\eqref{numberExcitationsE},
\begin{equation}
\label{E2PEFmeas}
    F_E = \frac{1}{2} \sigma_E \mathcal{T} Q n \int\limits_{-z_R}^{z_R}K(z)^{'} dz \int\limits_{\lambda_i}^{\lambda_f}\gamma(\lambda)\Phi(\lambda) d\lambda.
\end{equation}
To place an upper bound on the E2PA cross-section we replace $F_{E}$ in Eq.~\eqref{E2PEFmeas} with the measurable fluorescence lower bound $F^{\mathrm{LB}}$ (cnt~s$^{-1}$) and solve for $\sigma_{E}$, which becomes the cross-section upper bound, $\sigma^{\mathrm{UB}}_{E}$ (cm$^2$~fluorophore$^{-1}$), 
\begin{equation}
\label{E2PAUB}
    \sigma^{\mathrm{UB}}_{E}  = \frac{2 F^{\mathrm{LB}}}{\mathcal{T} Q n \int\limits_{-z_R}^{z_R}K(z)^{'} dz \int\limits_{\lambda_i}^{\lambda_f}\gamma(\lambda)\Phi(\lambda) d\lambda}.
\end{equation}
$F^{\mathrm{LB}}$ is measured (Appendix~\ref{DataAcq}), $\mathcal{T}$ is calculated based on the manufacturer's specifications, $Q$ is measured (Appendix~\ref{MeanPhotNum}) and the parameters $n$, $K(z)^{'}$, $\gamma(\lambda)$ and $\Phi(\lambda)$ are found in the methods described for C2PEF. All parameters are listed in Tables~\ref{Tab:sampleParam} and~\ref{Tab:expParam}. 

In order to generate a curve for E2PEF as a function of the mean photon number (the diagonals in Fig.~\ref{fig:sensPlot}), the slope $\frac{F_E}{Q/g}$ is solved for in Eq.~\eqref{E2PEFmeas} and a selected $\sigma_{E}$ is used. This slope is multiplied by the mean photon number on the horizontal axes. 

The uncertainty on our C2PA cross-sections and E2PA cross-section upper bounds are calculated by propagating the errors in all the measured and calculated parameters that go into either Eq.~\eqref{C2PAcrosssection} or Eq.~\eqref{E2PAUB}. We multiply these values by the coverage factor ($k = 2$, see Appendix~\ref{DataAcq}). The uncertainty in C2PA cross-sections is then $\approx 28\%$ and for E2PA cross-section upper bounds $\approx 24\%$.

\providecommand{\noopsort}[1]{}\providecommand{\singleletter}[1]{#1}%

\end{document}